\documentclass[nofootinbib,prd,aps,superscriptaddress,preprintnumbers,preprint]{revtex4}

\pdfoutput=1
\usepackage[english]{babel}
\usepackage{amsmath,amssymb,bm,slashed}
\usepackage{multirow}
\usepackage{graphicx}
\usepackage[sort&compress]{natbib}
\usepackage{xcolor}
\usepackage[normalem]{ulem}
\definecolor{red}{rgb}{1.0, 0, 0}

\allowdisplaybreaks

\bibpunct{[}{]}{,}{n}{}{,}

\newcommand{\bra}[1]{\ensuremath{\langle #1 |}}   
\newcommand{\ket}[1]{\ensuremath{| #1 \rangle}}   
\newcommand{\bigbra}[1]{\ensuremath{\big\langle #1 \big|}}   
\newcommand{\bigket}[1]{\ensuremath{\big| #1 \big\rangle}}   

\newcommand{\BR}{\text{BR}}

\renewcommand{\vec}[1]{{\mathbf{#1}}}
\renewcommand{\Re}{{\text{Re}}}
\renewcommand{\Im}{{\text{Im}}}

\begin{document}

\title{Flavor Violating Higgs Decays}
\author{Roni Harnik}             \email[Email: ]{roni@fnal.gov}
\affiliation{Fermilab, P.O.~Box 500, Batavia, IL 60510, USA}
\author{Joachim Kopp}            \email[Email: ]{jkopp@fnal.gov}
\affiliation{Fermilab, P.O.~Box 500, Batavia, IL 60510, USA}
\affiliation{Max Planck Institute for Nuclear Physics, PO Box 103980, 69029 Heidelberg, Germany}
\author{Jure Zupan}              \email[Email: ]{zupanje@ucmail.uc.edu}
\affiliation{Department of Physics, University of Cincinnati, Cincinnati, Ohio 45221,USA}
\date{September 10, 2012} 

\begin{abstract}
  We study a class of nonstandard interactions of the newly discovered 125~GeV
  Higgs-like resonance that are especially interesting probes of new physics:
  flavor violating Higgs couplings to leptons and quarks. These interaction can arise in
  many frameworks of new physics at the electroweak scale such as two 
  Higgs doublet models, extra dimensions, or models of compositeness.
  We rederive  constraints on flavor violating Higgs couplings using data on rare decays,
  electric and magnetic dipole moments, and meson oscillations. We confirm that
  flavor violating Higgs boson decays to leptons can be sizeable with, e.g.,
  $h \to \tau\mu$ and $h\to \tau e$ branching ratios  of ${\mathcal O}( 10\%)$
  perfectly allowed by low energy constraints. We estimate the current LHC
  limits on  $h\to \tau\mu$ and $h \to \tau e$  decays by recasting existing
  searches for the SM Higgs in the $\tau\tau$ channel and find that these
  bounds are already stronger than those from rare tau decays. We also show
  that these limits can be improved significantly with dedicated searches and
  we outline a possible search strategy.  Flavor violating Higgs decays
  therefore present an opportunity for discovery of new physics which in some
  cases may be easier to access experimentally than flavor conserving
  deviations from the Standard Model Higgs framework.
\end{abstract}

\begin{flushright}
  FERMILAB-PUB-12-498-T
\end{flushright}

\maketitle

\section{Introduction}
\label{sec:introduction}

Both ATLAS and CMS have recently announced  the discovery of a Higgs-like resonance
with a mass of  $m_h\simeq 125$~GeV
\cite{ATLAS-CONF-2012-093, CMS-PAS-HIG-12-020, ATLASHiggs, CMSHiggs}, further
supported by combined Tevatron data~\cite{Aaltonen:2012qt}.  An interesting
question is whether the properties of this resonance are consistent with the
Standard Model (SM) Higgs boson. Deviations from the SM predictions could point to
the existence of a secondary mechanism of electroweak symmetry breaking or to
other types of new physics not too far above the electroweak scale.  While
there is a large ongoing experimental effort to measure precisely the decay
rates into the channels that dominate for the SM Higgs, it is equally important
to search for Higgs decays into channels that are subdominant or absent in the
SM. For instance, since the couplings of the Higgs boson to quarks of the first
two generations and to leptons are suppressed by small Yukawa couplings in the
SM, new physics contributions can easily dominate over the SM predictions.
Another possibility, and the main topic of this paper, is
\emph{flavor violating} (FV) Higgs decays, for instance into $\tau\mu$ or $\mu
e$ final states. The study of FV couplings of the Higgs boson has a long
history \cite{Bjorken:1977vt, McWilliams:1980kj, Shanker:1981mj, Barr:1990vd,
Babu:1999me, DiazCruz:1999xe, Han:2000jz, Blanke:2008zb, Casagrande:2008hr,
Giudice:2008uua, AguilarSaavedra:2009mx, Albrecht:2009xr, Buras:2009ka,
Agashe:2009di, Goudelis:2011un, Arhrib:2012mg, McKeen:2012av, Azatov:2009na,
Blankenburg:2012ex, Kanemura:2005hr, Davidson:2010xv}. In this paper, we refine the
indirect bounds on the FV couplings. Most importantly,
we discuss in detail  possible search strategies for FV Higgs decays
at the LHC and derive for the first time limits from LHC data. 

As pointed out in the previous literature, and confirmed by the present
analysis, the indirect constraints on many FV Higgs decays are rather weak.
In particular, the branching ratios for $h \to \tau\mu$ and $h \to \tau e$ can
reach up to 10\%~\cite{Blankenburg:2012ex}.
In fact, for $h\to \tau\mu$ and $h\to \tau e$  already now\footnote{By $h\to \tau\mu$ we
always mean the sum of $h\to \tau^+\mu^-$ and $h\to \tau^-\mu^+$ and similarly
for the other decay modes.}, without targeted searches, the LHC is
placing limits that are comparable to or even stronger than those from rare
$\tau$ decays.  As we shall see later, re-casting a $h\to\tau\tau$ analysis
with 4.7~fb$^{-1}$ of 7~TeV ATLAS data~\cite{Chatrchyan:2012vp} gives a bound on
the branching fraction of the Higgs into $\tau\mu$ or $\tau e$ around 10\%. We
will also demonstrate that dedicated searches can be much more sensitive.

These decays could thus give a striking signature of new physics at the LHC, and we
strongly encourage our experimental colleagues to include them in their
searches. Another experimentally interesting set of decay channels are flavor
conserving decays to the first two generations, e.g., $h\to \mu^+\mu^-$, on
which we will comment further below.  We emphasize that large deviations from
the SM do not require very exotic flavor structures. A branching ratio for
$h\to \tau\mu$ comparable to the one for $h\to \tau\tau$, or a $h\to
\mu^+\mu^-$ branching ratio a few times larger than in the SM
can arise in many models of flavor (for instance in models with
continuous and/or discrete flavor symmetries \cite{Ishimori:2010au}, or in
Randall-Sundrum models \cite{Perez:2008ee}) as long as there is new physics at the
electroweak scale and not just the SM.  The lepton flavor violating decay $h
\to \tau\mu$ has been studied in~\cite{DiazCruz:1999xe}, and it was found that
the branching ratio for this decay can be up to 10\% in certain Two Higgs
Doublet Models (2HDMs).

In fact, there may already be experimental hints that the Higgs couplings to
fermions may not be SM-like. For instance, the BaBar collaboration recently
announced a $3.4\sigma$ indication of flavor universality violation in $b\to
c\tau\nu$ transitions~\cite{Lees:2012xj}, which can be explained for instance by
an extended Higgs sector with nontrivial flavor structure~\cite{Fajfer:2012jt}.

The paper is organized as follows. In Sec.~\ref{sec:framework} we introduce the
theoretical framework we will use to parameterize the flavor violating decays of
the Higgs. In Sec.~\ref{sec:leptons} we derive bounds on flavor violating Higgs
couplings to leptons and translate these bounds into limits on the Higgs decay
branching fractions to the various flavor violating final states. In
Sec.~\ref{sec:quarks} we do the same for flavor violating couplings to quarks.
We shall see that decays of the Higgs to $\tau\mu$ and to $\tau e$ with
sizeable branching fractions are allowed, and that also flavor violating
couplings of the Higgs to top quarks are only weakly constrained.  Motivated by
this we turn to the LHC in Section~\ref{sec:lhc} and estimate the current
bounds on Higgs decays to $\tau\mu$ and $\tau e$ using data from an existing $h
\to \tau\tau$ search.  We also discuss a strategy for a dedicated $h\to\tau\mu$
search and comment on differences with the SM $h\to\tau\tau$ searches. We will
see that the LHC can make significant further progress in probing the Higgs'
flavor violating parameters space with existing data. We conclude in
Section~\ref{sec:conclusions}.  In the appendices, we give more details on the
calculation of constraints from low-energy observables.

\section{The framework}
\label{sec:framework}

After electroweak symmetry breaking (EWSB) the fermionic mass terms and the
couplings of the Higgs boson to fermion pairs in the mass basis are in general
\begin{align}
  {\cal L}_Y = - m_i \bar f_L^i f_R^i - Y_{ij}(\bar f_L^i f_R^j) h + h.c. + \cdots \,,
  \label{eq:Yukawa}
\end{align}
where ellipses denote nonrenormalizable couplings involving more than one Higgs
field operator. In our notation, $f_L=q_L, \ell_L$ are $SU(2)_L$ doublets,
$f_R=u_R, d_R, \nu_R, \ell_R$ the weak singlets, and indices run over
generations and fermion flavors (quarks and leptons) with summation implicitly
understood.  In the SM the Higgs couplings are diagonal,
$Y_{ij}=(m_i/v)\delta_{ij}$, but in general NP models the structure of the
$Y_{ij}$ can be very different. Note that we use the normalization $v=246$ GeV
here.  The goal of the paper is to set bounds on $Y_{ij}$ and identify
interesting channels for Higgs decays at the LHC.  Throughout we will assume
that the Higgs is the only additional degree of freedom with mass ${\mathcal
O}(\text{100~GeV})$ and that the $Y_{ij}$'s are the only source of flavor violation.
These assumptions are not necessarily valid in general, but will be a good
approximation in many important classes of new physics frameworks.  Let us now show
how $Y_{ij} \neq (m_i/v)\delta_{ij}$ can arise in two qualitatively different
categories of NP models.

\paragraph{A single Higgs theory.}
Let us first explore the possibility that the Higgs is the only field that
causes EWSB (see
also~\cite{Giudice:2008uua,Azatov:2009na,Agashe:2009di,Buchmuller:1985jz,delAguila:2000aa,delAguila:2000rc,Babu:1999me}).  For simplicity let us also assume that at energies below $\sim
200$~GeV the spectrum consists solely of the SM particles: three generations of
quarks and leptons, the SM gauge bosons and the Higgs at 125~GeV.  Additional
heavy fields (e.g.\ scalar or fermionic partners which address the hierarchy
problem) can be integrated out, so that we can work in effective field theory
(EFT)---the effective Standard Model.  In addition to the SM Lagrangian
\begin{align}
  \begin{split}
    {\cal L}_{SM} &= \bar f^j_L i\slashed D f^j_L + \bar f^j_R i\slashed D f_R^j
      -\big[ \lambda_{ij} (\bar f_L^i f_R^j) H + h.c. \big]
      + D_\mu H^\dagger D^\mu H - \lambda_H\Big(H^\dagger H - \tfrac{v^2}{2}\Big)^2 \,,
  \end{split}
\end{align}
there are then also higher dimensional terms due to the heavy degrees of
freedom that were integrated out:
\begin{align}
  \Delta {\cal L}_Y &= -\frac{\lambda'_{ij}}{\Lambda^2} (\bar f_L^i f_R^j) H (H^\dagger H)
                         + h.c. + \cdots \,,
  \label{eq:eff:lagr}
\end{align}
Here we have written out explicitly only the terms that modify the Yukawa
interactions. We can truncate the expansion  after the terms of dimension~6,
since these already suffice to completely decouple the values of the fermion
masses from the values of fermion couplings to the Higgs boson.  Additional dimension~6 operators involving
derivatives include
\begin{align}
  \begin{split}
    \Delta {\cal L}_D &= \frac{\lambda_L^{ij}}{\Lambda^2} (\bar f^i_L\gamma^\mu  f^j_L)
                           (H^\dagger i \overleftrightarrow{D_\mu} H)
                       + \frac{\lambda_R^{ij}}{\Lambda^2} (\bar f^i_R\gamma^\mu  f^j_R)
                           (H^\dagger i \overleftrightarrow{D_\mu} H) + \cdots \,,
    \label{eq:extra-dim6-terms}
  \end{split}
\end{align}
where $(H^\dagger i \overleftrightarrow{D_\mu} H)\equiv H^\dagger i {D_\mu} H-
(i {D_\mu} H^\dagger) H$.  The couplings $\lambda_{ij}'$ are complex in
general, while the $\lambda_{L,R}^{ij}$ are real.  The derivative couplings do not
give rise to fermion-fermion-Higgs couplings after EWSB and are irrelevant for
our analysis.  In Eq.~\eqref{eq:extra-dim6-terms} there are in principle also
terms of the form $(\bar f^i_{L,R}i \slashed D f^j_{L.R}) H^\dagger H $, which,
however, can be shown to be equivalent to \eqref{eq:eff:lagr} by using
equations of motion.

After electroweak symmetry breaking (EWSB) and diagonalization of the mass
matrices, one obtains the Yukawa Lagrangian in Eq. \eqref{eq:Yukawa}, with
\begin{align}
  \sqrt{2} m = V_{L}  \bigg[ \lambda + \frac{v^2}{2\Lambda^2} \lambda' \bigg] V_{R}^\dagger\, v
  \label{eq:mass}\,,\qquad
  \sqrt{2} Y = V_{L}  \bigg[ \lambda + 3 \frac{v^2}{2\Lambda^2}\lambda' \bigg]  V_{R}^\dagger\,,
 \end{align}
where the unitary matrices $V_{L}, V_{R}$  are those which diagonalize the mass
matrix, and $v=246$~GeV.  In the mass basis we can write
\begin{align}
  Y_{ij} = \frac{m_i}{v}\delta_{ij} + \frac{v^2}{\sqrt2 \Lambda^2} \hat \lambda_{ij}\,,
\end{align}
where $\hat \lambda= V_L \lambda' V_R$. 
In the limit $\Lambda\to \infty$ one obtains the SM, where
the Yukawa matrix $Y$ is diagonal, $Y v=m$. For $\Lambda$ of the order of the
electroweak scale, on the other hand, the mass matrix and the couplings of the
Higgs to fermions can be very different as $\hat \lambda$ is in principle an
arbitrary non-diagonal matrix.

Taking the off diagonal Yukawa couplings nonzero can come with a theoretical
price.  Consider, for instance, a two flavor mass matrix involving $\tau$ and
$\mu$.  If the off-diagonal entries are very large the mass spectrum is
generically not hierarchical. A hierarchical spectrum would require a delicate
cancellation among the various terms in Eq.~(\ref{eq:mass}).  Tuning is
avoided if \cite{Cheng:1987rs}
\begin{align}
  |Y_{\tau\mu}Y_{\mu\tau}| \lesssim \frac{m_\mu m_\tau}{v^2}\,,
\end{align}
with similar conditions for the other off diagonal elements.  Even though we
will keep this condition in the back of our minds, we will not restrict the
parameter space to fulfill it.

\paragraph{Models with several sources of EWSB:}
Let us now discuss the case where the Higgs at 125~GeV is not the only scalar
that breaks electroweak symmetry. The modification of the above discussion is
straightforward. The additional sources of EWSB are assumed to be heavy and can
thus still be integrated out. Their EWSB effects can be described by a spurion
$\chi$ that formally transforms under electroweak global symmetry and then
obtains a vacuum expectation value (vev), which breaks the electroweak
symmetry. If $\chi$ has the quantum numbers $(2,1/2)$ under $SU(2)_L\times
U(1)_Y$ it can contribute to  quark and lepton masses.\footnote{A spurion which
transforms as a triplet can also contribute to Majorana masses for neutrinos.}
This allows the Yukawa interactions $Y$ of the 125~GeV Higgs to be misaligned
with respect to the fermion mass matrix $m$  in Eq.~\eqref{eq:Yukawa}.

The simplest example for a full theory of this class is a type III two Higgs doublet
model (2HDM) where both Higgses obtain a vev and couple to fermions.
 In
the full theory both of the scalars then have a Lagrangian of the form
\eqref{eq:Yukawa}
\begin{align}
  {\cal L}_Y &= - m_i \bar f_L^i f_R^i - Y_{ij}^a (\bar f_L^i f_R^j) h^a + h.c. + \cdots,
\end{align}
where the index $a$ runs over all the scalars (with $Y^a_{ij}$ imaginary for
pseudoscalars), and $m_i$ receives contributions from both vevs. In addition
there is also a scalar potential which mixes the two Higgses. Diagonalizing the
Higgs mass matrix then also changes $Y_{ij}^a$, but removes the Higgs mixing.
For our purposes it is simplest to work in the Higgs mass basis. All the
results for a single Higgs are then trivially modified, replacing our final
expressions below by a sum over several Higgses. For a large mass gap, where
only one Higgs is light, the contributions from the heavier Higgs are power
suppressed, unless its flavor violating Yukawa couplings are  parametrically
larger than those of the light Higgs. The contributions from the heavy Higgs
correspond to the higher dimensional operators discussed in the previous
paragraph. This example can be trivially generalized to models with many Higgs
doublets.

We next derive constraints on flavor violating Higgs couplings and work out the
allowed branching fractions for flavor violation Higgs decays.  In placing the
bounds we will neglect the FV contributions of the remaining states in the full
theory. Our bounds thus apply barring cancellations with these other terms.

\section{Leptonic flavor violating Higgs decays}
\label{sec:leptons}

The FV decays $h\to e\mu, e\tau, \mu\tau$ arise at tree level from the assumed
flavor violating Yukawa interactions,~Eq. \eqref{eq:Yukawa}, where the relevant
terms are explicitly
\begin{align}
\begin{split}
  \mathcal{L}_Y \supset &-Y_{e\mu} \bar{e}_L \mu_R h
                      - Y_{\mu e} \bar\mu_L e_R h - Y_{e\tau} \bar e_L \tau_R h
                      - Y_{\tau e} \bar\tau_L e_R h - Y_{\mu\tau} \bar\mu_L \tau_R h
                      - Y_{\tau\mu} \bar\tau_L \mu_R h
                      + h.c. \,.
\end{split}
\end{align}
The bounds on the FV Yukawa couplings are collected in Table \ref{tab:leptons}, where
for simplicity of presentation the flavor diagonal muon and tau Yukawa couplings,
\begin{align}
  \mathcal{L}_Y \supset -Y_{\mu\mu} \bar\mu_L \mu_R h - Y_{\tau\tau} \bar\tau_L \tau_R h +h.c.\,,
\end{align}
were set equal to their respective SM values $\big(Y_{\mu\mu}\big)_{\rm
SM}=m_\mu/v$, $\big(Y_{\tau\tau}\big)_{\rm SM}=m_\tau/v$.  Similar bounds on FV
Higgs couplings to quarks are collected in Table \ref{tab:light-quarks}.
Similar constraints on flavor violating Higgs decays have been present recently
also in~\cite{Blankenburg:2012ex}.  While our results agree qualitatively with
previous ones, small numerical differences are expected because we avoid some
of the approximations made by previous authors. We also consider some
constraining processes not discussed before.

\begin{table}
  \centering
  \parbox{13cm}{
  \begin{ruledtabular}
  \begin{tabular}{@{\qquad}lcc@{\qquad}}
    Channel  &    Coupling                               & Bound        \\ \hline
    $\mu \to e\gamma$         & $\sqrt{|Y_{\mu e}|^2 + |Y_{e \mu}|^2}$ & $< 3.6 \times 10^{-6} $ \\
    $\mu \to 3e$              & $\sqrt{|Y_{\mu e}|^2 + |Y_{e \mu}|^2}$ & $\lesssim 3.1 \times 10^{-5}$ \\
    electron $g-2$            & $\Re(Y_{e \mu} Y_{\mu e})$             & $-0.019\dots 0.026$ \\
    electron EDM              & $|\Im(Y_{e \mu} Y_{\mu e})|$           & $<9.8 \times 10^{-8}$ \\
    $\mu \to e$ conversion    & $\sqrt{|Y_{\mu e}|^2 + |Y_{e \mu}|^2}$ & $< 1.2 \times 10^{-5} $ \\
    $M$-$\bar M$ oscillations & $|Y_{\mu e} + Y_{e\mu}^*|$             & $< 0.079$ \\
    \hline
    $\tau \to e\gamma$        & $\sqrt{|Y_{\tau e}|^2 + |Y_{e \tau}|^2}$ & $< 0.014$ \\
    $\tau \to 3 e$            & $\sqrt{|Y_{\tau e}|^2 + |Y_{e \tau}|^2}$ & $\lesssim 0.12$ \\
    electron $g-2$            & $\Re(Y_{e \tau} Y_{\tau e})$             & $[-2.1\dots 2.9] \times 10^{-3}$ \\
    electron  EDM             & $|\Im(Y_{e \tau} Y_{\tau e})|$           & $< 1.1 \times 10^{-8}$ \\
    \hline
    $\tau\to \mu\gamma$       & $\sqrt{|Y_{\tau\mu}|^2 + |Y_{\mu\tau}|^2}$ & $0.016$ \\
    $\tau\to 3\mu$            & $\sqrt{|Y_{\tau\mu}^2 + |Y_{\mu\tau}|^2} $ & $\lesssim 0.25$  \\
    muon $g-2$                & $\Re(Y_{\mu\tau} Y_{\tau\mu})$           & $(2.7\pm0.75) \times 10^{-3}$  \\
    muon EDM                  & $\Im(Y_{\mu\tau} Y_{\tau\mu})$           & $ -0.8 \dots 1.0$ \\
    \hline
    $\mu \to e \gamma$        & $\big(|Y_{\tau\mu}Y_{e \tau}|^2 + |Y_{\mu\tau}Y_{\tau e}|^2\big)^{1/4}$
                                                                         & $< 3.4 \times 10^{-4}$
  \end{tabular}
  \end{ruledtabular}}
  \caption{Constraints on flavor violating Higgs couplings to $e$, $\mu$,
  $\tau$ for a Higgs mass $m_h = 125$~GeV and assuming that the flavor diagonal
  Yukawa couplings equal the SM values (see text for details). For the muon magnetic
  dipole moment we show the value of the couplings required to explain the
  observed~$\Delta a_\mu$ (if this is used only as an upper bound one has
  $\sqrt{\Re(Y_{\mu\tau}Y_{\tau \mu})}<0.065$ at $95\%$CL).}
  \label{tab:leptons}
\end{table}

We first give more details on how the bounds in Tables~\ref{tab:leptons} and
\ref{tab:light-quarks} were obtained and then move on to predictions for
the allowed sizes of the FV Higgs decays.

\subsection{Constraints from ${\bm \tau \to \mu\gamma}$, ${\bm \tau \to e \gamma}$
            and ${\bm \mu \to e\gamma}$ }
\label{sec:tmg}

\begin{figure}
  \begin{center}
    \includegraphics[width=0.85\textwidth]{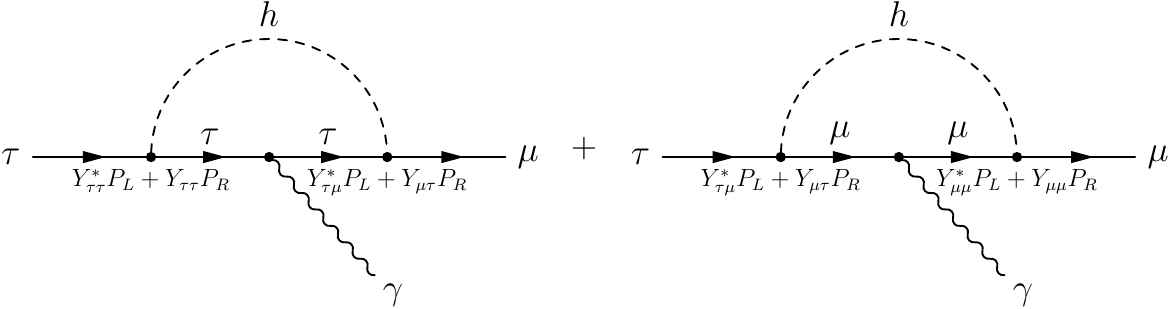}
  \end{center}
  \vspace{-10mm}
  \caption{Diagrams contributing to the flavor violating decay $\tau \to \mu\gamma$,
    mediated by a Higgs boson with flavor violating Yukawa couplings.}
  \label{fig:tau-mu-gamma}
\end{figure}

The effective Lagrangian  for the $\tau \to \mu\gamma$ decay is given by
\begin{align}
  {\cal L}_{\rm eff} &= c_L  Q_{L\gamma} + c_R Q_{R\gamma} +h.c. \,,
  \label{eq:O-tau-mu-gamma}
\end{align}
where the dim-5 electromagnetic penguin operators are
\begin{align}\label{QEMP}
  Q_{L\gamma, R\gamma} &= \frac{e}{8\pi^2} m_\tau
    \big(\bar\mu\,\sigma^{\alpha\beta} P_{L,R} \tau\big) F_{\alpha\beta} \,,
\end{align}
with $\alpha, \beta$ the Lorentz indices and $F_{\alpha\beta}$ the
electromagnetic field strength tensor. The Wilson coefficients $c_L$ and $c_R$
receive contributions from the two 1-loop diagrams shown in
Fig.~\ref{fig:tau-mu-gamma} (with the first one dominant), and a comparable
contribution from Barr-Zee type 2-loop diagrams, see Fig.~\ref{fig:2loop} in
Appendix~\ref{Appendix:FCNC}.  The complete one loop and two loop expressions
are given in Appendix \ref{Appendix:FCNC}.

In the approximation $Y_{\mu\mu} \ll Y_{\tau\tau}$, only the first of the one-loop
diagrams in Fig.~\ref{fig:tau-mu-gamma} is relevant (in addition to the 2-loop
diagrams).  Using also $m_\mu \ll m_\tau \ll m_h$ and assuming $Y_{\mu\mu}$,
$Y_{\tau\tau}$ to be real, the expressions for the one-loop Wilson coefficients
$c_L$ and $c_R$ simplify to (this agrees with \cite{Blankenburg:2012ex})
\begin{align}
  c_L^{\rm 1loop} \simeq \frac{1}{12 m_h^2} Y_{\tau\tau} Y_{\tau\mu}^*
    \bigg(\! -4 + 3 \log \frac{m_h^2}{m_\tau^2} \bigg) \,,\quad
  c_R^{\rm 1loop} \simeq \frac{1}{12  m_h^2} Y_{\mu\tau} Y_{\tau\tau}
    \bigg(\! -4 + 3 \log \frac{m_h^2}{m_\tau^2} \bigg) \,.
  \label{eq:tmg-CR-simplified}
\end{align}
The 2-loop contributions are numerically
\begin{align}
  c_L^{\rm 2loop} &= Y_{\tau \mu}^* (-0.082 \,Y_{tt}+0.11) \frac{1}{(125 {\rm GeV})^2}
                   = 0.055  Y_{\tau \mu}^* \frac{1}{(125 {\rm GeV})^2},
  \label{eq:cL:2loop}
\end{align}
where in the last step we used for the top Yukawa coupling
$Y_{tt}=(Y_{tt})_{SM}=\bar m_t/v=0.67$, and we have normalized the results to
$m_h$ for easier comparison. (By $\bar{m}_t$, we denote the top quark mass
parameter in the $\overline{\text{MS}}$ renormalization scheme, $\bar{m_t} \simeq 164$~GeV.) 
The analytical form of the Wilson
coefficient can be found in Appendix \ref{Appendix:FCNC}. The same result
applies to $c_R^{\rm 2loop}$ with the replacement $Y_{\tau\mu}^*\to
Y_{\mu\tau}$. The 2-loop contribution consists is dominated by two terms, the one with the top quark in the loop and of a somewhat larger $W$ contribution. They have opposite signs and thus part of $W$ contributions is cancelled. The end result has an increased
sensitivity to the precise value of $Y_{tt}$.  For $Y_{tt}\simeq \bar m_t/v$ the
2-loop contribution is about four times as large as the 1-loop contribution, while
for other values of $Y_{tt}$ (e.g., $Y_{tt}\simeq -m_t/v$) the 2-loop
contribution can be an order of magnitude larger. Note that we keep complete 2-loop expressions \cite{Chang:1993kw}, including the finite terms, while in \cite{Goudelis:2011un,Blankenburg:2012ex} only the leading log term of the top loop contribution was kept. Numerically, this amounts to an ${\mathcal O}(1)$ difference. 

In terms of the Wilson coefficients $c_L$ and $c_R$, the rate for $\tau \to
\mu\gamma$ is
\begin{align}
  \Gamma(\tau \to \mu\gamma) = \frac{\alpha \,m_\tau^5}{64\pi^4}
    \big( |c_L|^2 + |c_R|^2 \big) \,.
  \label{eq:Gamma-tmg}
\end{align}
Using a Higgs mass $m_h = 125$~GeV and assuming $Y_{\tau\tau} = m_\tau / v$,
$Y_{tt} = \bar{m}_t / v$, we can then translate the experimental bound $
\BR(\tau \to \mu\gamma) < 4.4 \times 10^{-8}$~\cite{Beringer:2012PDG} into a
constraint $\sqrt{|Y_{\tau\mu}|^2 + |Y_{\mu\tau}|^2} < 1.6 \times 10^{-2}$ (see
Table \ref{tab:leptons}). The bound is relaxed if $Y_{\tau\tau}$ and/or
$Y_{tt}$ are smaller than their SM values.

The expressions for $\mu\to e\gamma$ and $\tau \to e\gamma$ are obtained in an
analogous way with the obvious replacements ($\tau \to \mu, \mu\to e$ for the
first and $\mu\to e$ for the second in Eqs.~\eqref{QEMP},
\eqref{eq:tmg-CR-simplified}, \eqref{eq:cL:2loop}, \eqref{eq:Gamma-tmg}). The
experimental bound $  \BR(\mu \to e \gamma) < 2.4 \cdot 10^{-12}$
\cite{Beringer:2012PDG} then translates to $  \sqrt{|Y_{\mu e}|^2 +
|Y_{e\mu}|^2} < 3.6 \times 10^{-6}$ and $  \BR(\tau \to e \gamma) < 3.3\times
10^{-8}$  \cite{Beringer:2012PDG} to a constraint $  \sqrt{|Y_{\tau e}|^2 +
|Y_{e\tau}|^2} < 1.4 \times 10^{-2}$ using the SM values for $Y_{\tau\tau},
Y_{\mu\mu}, $ and $Y_{tt}$. The $\mu\to e\gamma$ bound is completely dominated
by the two loop contribution, while for $\tau\to e\gamma$, the two loop and one
loop contributions are comparable. 

The decay $\mu\to e\gamma$ can also be used to place a bound on the combination
$Y_{\mu\tau} Y_{\tau e}$ using the 1-loop Wilson coefficient (in agreement with \cite{Blankenburg:2012ex})
 \begin{align}
  c_L^{\rm 1loop} \simeq \frac{1}{8 m_h^2} \frac{m_\tau}{m_\mu}Y_{\mu\tau}^* Y_{\tau e}^*
    \bigg(\! -3 + 2 \log \frac{m_h^2}{m_\tau^2} \bigg) \,.
  \label{eq:tmg-CL-simplifiedmue}
\end{align}
As before, $c_R^{\rm 1loop}$ is obtained by replacing $Y_{\mu \tau}^*Y_{\tau
e}^*$ by $Y_{\tau \mu} Y_{e\tau}$. The 2-loop contribution is proportional to
$Y_{\mu e}$ and $Y_{e\mu}$. Setting them to zero, one obtains a bound
$\big(|Y_{\tau\mu}Y_{e \tau}|^2+ |Y_{\mu\tau}Y_{\tau e}|^2\big)^{1/4} < 3.4
\times 10^{-4} $.

\subsection{Constraints from $\bm \tau \to 3\mu$, $\bm\tau \to 3e$, $\bm\mu\to 3e$}

\begin{figure}
  \begin{center}
    \includegraphics[width=\textwidth]{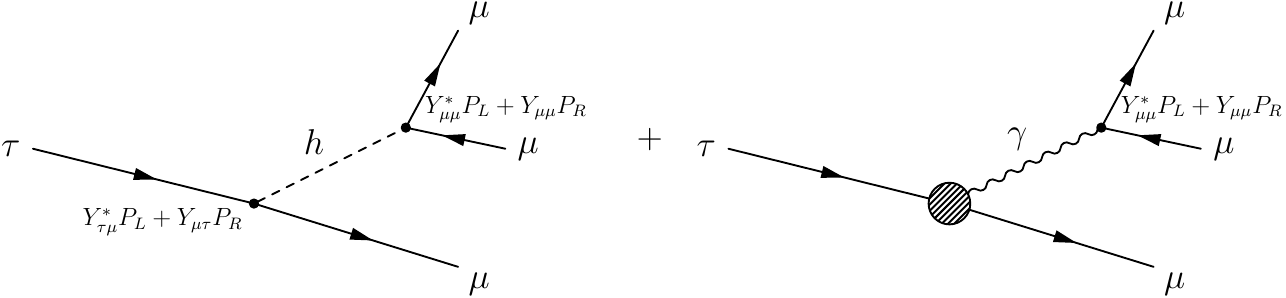}
    \vspace{-6mm}
  \end{center}
  \caption{Diagrams leading to  $\tau \to 3\mu$ decay. The tree level Higgs exchange
    contribution (left) is typically subdominant compared to higher-order
    contributions with the topology shown on the right. The blob represents
    loops of the form shown in Figs.~\ref{fig:tau-mu-gamma} and \ref{fig:2loop}.}
  \label{fig:tau-3mu}
\end{figure}

The decay $\tau \to 3\mu$ can be generated through tree level Higgs exchange, see the diagram in
Fig.~\ref{fig:tau-3mu} (left). However, the diagram is suppressed
not only by the flavor violating Yukawa couplings $Y_{\tau\mu}$ and $Y_{\mu\tau}$,
but also by the flavor-conserving coupling $Y_{\mu\mu}$. It is thus
subleading compared to the higher order contributions: the 1-loop diagrams
of the form shown in Fig.~\ref{fig:tau-mu-gamma} and 2-loop diagrams like the ones
shown in Fig.~\ref{fig:2loop} (for $\tau \to \mu\gamma$). These generate $\tau \to 3\mu$
if the outgoing gauge boson is off-shell and ``decays'' to a muon pair.
This general topology is shown in the right part of Fig.~\ref{fig:tau-3mu}.

Integrating out the Higgs, the heavy gauge bosons and the top quark, these
contributions match onto an effective Lagrangian. The full effective Lagrangian
is similar to the one in Eq.~\eqref{eq:mue:eff} for $\mu \to e$ conversion, but
with quarks replaced by muons. Since a full evaluation of the 2-loop
contributions is beyond the scope of this work, we will estimate the $\tau \to
3\mu$ rate by including only the dimension 5 elecromagnetic dipole
contributions of the form given in Eq.~\eqref{QEMP}. For $c_L$ and $c_R$, we
use the same expressions as for $\tau \to \mu\gamma$, see Sec.~\ref{sec:tmg}
and Appendix~\ref{Appendix:FCNC}. We evaluate these expressions at $q^2 = 0$.
We have checked that the neglected contributions are numerically smaller than
the dipole terms at one loop. At two loops, to the best of our knowledge, a
full evaluation of all potentially relevant diagrams is not available.

The corresponding expression for the flavor violating partial width of the $\tau$
is
\begin{align}
  \Gamma(\tau \to 3\mu) &\simeq \frac{\alpha^2 m_\tau^5}{6 (2\pi)^5} \,
  \bigg| \log\frac{m_\mu^2}{m_\tau^2} - \frac{11}{4} \bigg| \,
    \big(|c_L|^2 + |c_R|^2\big) \,,
  \label{eq:t3m-new}
\end{align}
where we have neglected terms additionally suppressed by the muon mass.  The
Wilson coefficients $c_L$ and $c_R$ are given approximately by
Eqs.~\eqref{eq:tmg-CR-simplified} and \eqref{eq:cL:2loop}, with the 2-loop
contribution dominating over the 1-loop one.  In addition to \eqref{eq:t3m-new}
there are also contributions to the $\tau\to 3\mu$ rate from effective flavor
violating $Z$ vertices induced at 1-loop by flavor violating Higgs exchanges.
These have the same scaling in terms of masses and the Yukawas as
\eqref{eq:t3m-new}, but are found to be numerically an order of magnitude
smaller \cite{Goto:2015iha}. We therefore neglect them in the following.

The experimental bound $\BR(\tau \to 3\mu) < 2.1 \times
10^{-8}$~\cite{Nakamura:2010zzi} translates into a constraint
$\sqrt{|Y_{\tau\mu}^2 + |Y_{\mu\tau}|^2} < 0.25$ for $m_h = 125$~GeV and
assuming that the diagonal Yukawa couplings $Y_{\tau\tau}$, $Y_{\mu\mu}$ and
$Y_{tt}$ have their Standard Model values.  The decay $\tau \to 3\mu$ thus
leads to a weaker limit on $Y_{\tau\mu}$, $Y_{\mu\tau}$ than $\tau \to
\mu\gamma$, mainly because $\Gamma(\tau \to 3\mu)$ is suppressed by an
additional power of $\alpha$ compared to $\Gamma(\tau \to \mu\gamma$).

Similarly, the constraints on $Y_{\mu e}$, $Y_{e\mu}$, $Y_{\tau e}$,
$Y_{e\tau}$ following from the processes $\mu\to 3e$ and $\tau \to 3e$
are weaker than the corresponding limits from $\mu \to e\gamma$ and $\tau \to
e\gamma$. The bounds in Table \ref{tab:leptons} are obtained using the
experimental results $\BR(\mu \to 3e) < 1.0 \times
10^{-12}$~\cite{Bellgardt:1987du} and $\BR(\tau \to 3e) < 2.7 \times
10^{-8}$~\cite{Hayasaka:2010np}. We have also considered the process
$\tau \to e\mu\mu$, but found that it yields a weaker limit than $\tau
\to 3e$, mainly because of the smaller phase space.

\subsection{Constraints from muonium--antimuonium oscillations}

\begin{figure}
  \begin{center}
    \includegraphics[width=0.35\textwidth]{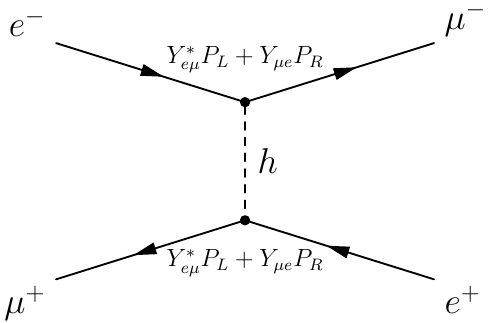}
  \end{center}
  \vspace{-9mm}
  \caption{Diagram leading to muonium--antimuonium oscillations.}
  \label{fig:MMbar}
\end{figure}

A $\mu^+ e^-$ bound state (called muonium $M$) can oscillate into an $e^+
\mu^-$ bound state (antimuonium $\bar{M}$) through the diagram in
Fig.~\ref{fig:MMbar}. The time-integrated $M\to \bar M$ conversion probability
is constrained by the MACS experiment at PSI~\cite{Willmann:1998gd} to be below
$P(M \to \bar{M}) < 8.3 \times 10^{-11} / S_B$, where the correction factor
$S_B\leq 1$ accounts for the splitting of muonium states in the magnetic field
of the detector. It depends on the Lorentz structure of the conversion operator
and varies between $S_B=0.35$ for $(S\pm P) \times (S\pm P)$ operators and
$S_B=0.9$ for $P\times P$ operators~\cite{Willmann:1998gd}. Conservatively, we
use the smallest value $S_B=0.35$ throughout.  Since we will find that
$M$--$\bar{M}$ oscillation constraints are much weaker than those from from
$\mu \to e\gamma$ and $\mu \to e$ conversion, this approximation suffices for
illustrative purposes.

The theoretical prediction for the $M \to \bar{M}$ conversion rate is governed
by the mixing matrix element (see, e.g.,~\cite{Clark:2003tv})
\begin{align}
  \mathcal{M}_{\bar{M} M} =
    \bigbra{\uparrow_\mu \downarrow_{\bar{e}} - \downarrow_\mu \uparrow_{\bar{e}}} \,
    \frac{ \big[\bar{\mu} (Y_{e \mu}^* P_L + Y_{\mu e} P_R) e\big]
    \big[\bar{\mu} (Y_{e \mu}^* P_L + Y_{\mu e} P_R) e\big] }{2 m_h^2}\,
    \bigket{\uparrow_e \downarrow_{\bar{\mu}} - \downarrow_e \uparrow_{\bar{\mu}}} \,,
  \label{eq:M-MbarM}
\end{align}
where  $\uparrow_X$ and $\downarrow_X$ are the spin orientations of particle
$X$.  We can work in the non-relativistic limit here. For a contact
interaction, the spatial wave function of muonium, $\phi_{1s} = \exp(-r / a_M)
/ [\pi a_M^3]^{1/2}$, only needs to be evaluated at the origin.  (Here $r$ is
the electron--antimuon distance and $a_M = (m_e + m_\mu) / (m_e m_\mu \alpha)$
is the muonium Bohr radius.) The resulting mass splitting between the two mass
eigenstates of the mixed $M$--$\bar{M}$ system is~\cite{Clark:2003tv},
\begin{align}
  \Delta M = 2 \, |\mathcal{M}_{\bar{M} M}|
           = \frac{|Y_{\mu e} + Y_{e\mu}^*|^2}{2 \pi a^3 m_h^2},
  \label{eq:DeltaM-MMbar}
\end{align}
and the time-integrated conversion probability is
\begin{align}
  P(M \to \bar{M}) = \int_0^\infty\! dt\, \Gamma_\mu\,\sin^2 (\Delta M \, t) \, e^{-\Gamma_\mu t}
                   = \frac{2}{\Gamma_\mu^2 / (\Delta M)^2 + 4} \,.
  \label{eq:P-MMbar}
\end{align}
The bound from the MACS experiment~\cite{Willmann:1998gd} then translates into
$|Y_{\mu e} + Y_{e\mu}^*| < 0.079$.

\subsection{Constraints from magnetic dipole moments}

\begin{figure}
  \begin{center}
    \includegraphics[width=0.4\textwidth]{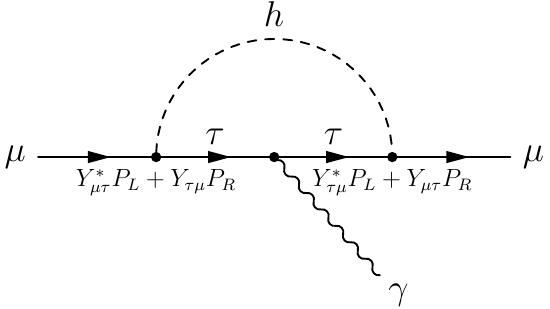}
  \end{center}
  \caption{A diagram contributing to the anomalous magnetic moment $g-2$ of the muon
  through FV couplings of the Higgs to $\tau\mu$.}
  \label{fig:g-2}
\end{figure}

The CP conserving and CP violating parts of the diagram in Fig.~\ref{fig:g-2}
generate magnetic and electric dipole moments of the muon, respectively.  Since
the experimental value of the magnetic dipole moment, $g_\mu-2$,  is above the
SM prediction at more than $3\sigma$, also the preferred value for the flavor
violating Higgs couplings will be nonzero.

The FV contribution to $(g-2)_\mu$ due to the $\tau$-Higgs loop in
Fig.~\ref{fig:g-2} is (neglecting  terms suppressed by $m_\mu / m_\tau$ or
$m_\tau / m_h$)
\begin{align}
  a_\mu \equiv \frac{g_\mu - 2}{2}
  &\simeq \frac{\Re(Y_{\mu\tau} Y_{\tau\mu})}{8\pi^2}
          \frac{m_\mu m_\tau}{2 m_h^2}
          \Big( 2 \log\frac{m_h^2}{m_\tau^2} - 3 \Big) \,,
  \label{eq:g-2}
\end{align}
in agreement with \cite{Blankenburg:2012ex}. The discrepancy between the measured value of $a_\mu$ and the one predicted by
the Standard Model~\cite{Nakamura:2010zzi,Bennett:2006fi},
\begin{align}
  \Delta a_\mu \equiv a_\mu^{\rm exp} - a_\mu^{\rm SM}
               = (2.87 \pm 0.63 \pm 0.49) \times 10^{-9},
  \label{eq:amu-discrepancy}
\end{align}
could thus be explained if there are FV Higgs interactions of the size
\begin{align}
  \Re(Y_{\mu\tau} Y_{\tau\mu}) \simeq (2.7 \pm 0.75) \times 10^{-3} \,,
  \label{eq:Ytm-g-2}
\end{align}
(for the definition of the Yukawa couplings see Eq.~\eqref{eq:Yukawa}).  This explanation
of $\Delta a_\mu$ requires $Y_{\mu\tau} \sim Y_{\tau\mu}$ to be a factor of a
few bigger than the SM value of the diagonal Yukawa, $m_\tau/v$, and is in
tension with limits from $\tau\to\mu\gamma$.\footnote{If the two loop
contribution to $\tau\to\mu\gamma$ is suppressed, e.g.\ due to a modification
of the top Yukawa coupling, which could lead to significant cancellation
between the 2-loop top and $W$ diagrams, there is a small region of parameter
space in which flavor violating Higgs couplings could explain the $(g-2)_\mu$
discrepancy without being ruled out by the one loop $\tau\to \mu\gamma$
constraint. We will, however, see below that even this case is disfavored
by the LHC limit derived in this paper (see Sec.~\ref{sec:lhc-current}).} It
is in further tension with the LHC limit extracted in Sec.~\ref{sec:lhc} of
this paper.

The measured $\Delta a_\mu$ could in principle also be explained by an enhanced
flavor conserving coupling of the muon to the Higgs if $Y_{\mu\mu} \sim 0.15
\sim 280 \, m_\mu/v$. However, in this case $h\to \mu\mu$ decays would be
enhanced to a level that is already ruled out by the searches at the LHC: From
the search for the MSSM neutral Higgs boson one obtains a bound $\sigma(gg\to
h\to \mu\mu) \lesssim 30\times  \sigma(gg\to h\to \mu\mu)_{\rm SM}$ or
$Y_{\mu\mu}\lesssim 5.5 m_\mu/v$~\cite{ATLAS-CONF-2012-094}.

\subsection{Constraints from electric dipole moments}

If the flavor violating Yukawa couplings in Fig.~\ref{fig:g-2} are complex, the
diagram shown there generates also an \emph{electric} dipole moment (EDM) for
the muon. The relevant term in the effective Lagrangian is
\begin{align}
  \mathcal{L}_{\rm EDM} = -\frac{i}{2} \, d_\mu \,\big( \bar{\mu} \sigma^{\alpha\beta}
    \gamma^5 \mu \big)F_{\alpha\beta}  \,,
  \label{eq:L-EDM}
\end{align}
with the electric dipole moment given by (neglecting the terms suppressed by
$m_\mu / m_\tau$ or $m_\tau / m_h$)
\begin{align}
  d_\mu \simeq -\frac{\Im(Y_{\mu\tau} Y_{\tau\mu})}{16\pi^2}
          \frac{e \, m_\tau}{2 m_h^2}
          \Big( 2 \log\frac{m_h^2}{m_\tau^2} - 3 \Big) \,,
  \label{eq:dmu}
\end{align}
in agreement with \cite{Blankenburg:2012ex}.
The experimental constraint $-10 \times 10^{-20} \, e\, \text{cm} < d_\mu < 8
\times 10^{-20} \, e \, \text{cm}$~\cite{Beringer:2012PDG} translates into the
rather weak limit $-0.8 \lesssim \Im(Y_{\mu\tau} Y_{\tau\mu}) \lesssim 1.0$.

A similar diagram with electrons instead of muons on the external legs also
contributes to the electron EDM, $d_e$. The experimental constraint
$|d_e|<0.105 \times 10^{-26} e$~cm~\cite{Beringer:2012PDG} translates into
$|\Im(Y_{e \tau} Y_{\tau e})|< 1.1 \times 10^{-8}$ for a tau running in the
loop, and into  $|\Im(Y_{e \mu} Y_{\mu e})| <9.8 \times 10^{-8}$ for a muon
running in the loop.

\subsection{Constraints from $\mu \to e$ conversion in nuclei}

\begin{figure}
  \begin{center}
    \includegraphics[width=\textwidth]{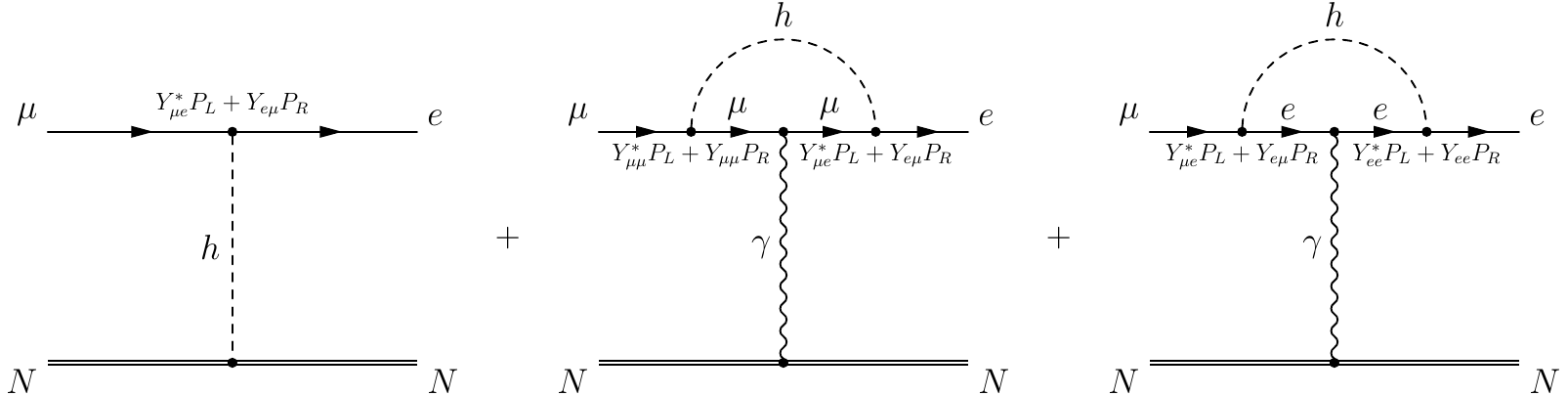}
  \end{center}
  \caption{Tree-level and one-loop diagrams contributing to $\mu\to e$ conversion
    in nuclei via the flavor violating
    Higgs Yukawa couplings $Y_{\mu e}$ and $Y_{e\mu}$.
    In addition, we also include numerically important two-loop diagrams,
    see Appendix~\ref{App:mutoe} for details.}
  \label{fig:mu-e-conversion}
\end{figure}

Very stringent constraints on the FV Yukawa couplings $Y_{\mu e}$ and
$Y_{e\mu}$ come from experimental searches for $\mu\to e$ conversion in
nuclei. The relevant tree-level and one-loop diagrams with one insertion of the FV Yukawa coupling are
shown in Fig.~\ref{fig:mu-e-conversion}. An effective scalar interaction arises
already at tree level from the first diagram in Fig.~\ref{fig:mu-e-conversion},
while vector and electromagnetic dipole
contributions arise at one loop level. 
We give complete expressions for the
tree level and one loop contributions in Appendix~\ref{App:mutoe}.
There are
also two-loop contributions, similar to the ones discussed in
Sec.~\ref{sec:tmg} in the context of $\mu \to e\gamma$. Numerically, the
two-loop contributions are larger than the one loop ones because they are not
suppressed by the small $Y_{\mu\mu}$ coupling but only by $Y_{tt}$ or the weak
gauge coupling. They are in fact comparbale to the
tree level contribution. Here, we always assume the diagonal
Yukawa couplings to have their SM values. With this assumption, the tree level
term is very sensitive to the strangeness content of the nucleon.

The bounds on the Yukawa couplings $Y_{e\mu}$ and $Y_{\mu e}$ from $\mu \to e$
conversion in nuclei, including tree level, one-loop and two-loop contributions,
are listed in Table \ref{tab:leptons}.

One could potentially also obtain interesting limits on $|Y_{e\tau}|$ and
$|Y_{\tau e}|$ from $\mu \to e$ conversion in nuclei, even though this requires
diagrams proportional to \emph{two} FV Yukawa couplings, because the other
constraints on these couplings are weak. The combinations $Y_{e\tau} Y_{\tau
\mu}$,  $Y_{e\tau} Y_{\mu\tau}^*$, $Y_{\tau e}^* Y_{\tau\mu}$ and $Y_{\tau e}^*
Y_{\mu \tau}^*$ are constrained by $\mu\to e$ conversion through 1-loop
diagrams similar to the ones shown in Fig.~\ref{fig:mu-e-conversion}, but with
a $\tau$ running in the loop (see Eq.~\eqref{eq:mue-G}). In the simplest case,
$Y_{e\tau} = Y_{\tau e}$, $Y_{\mu\tau} = Y_{\tau\mu}$, with all Yukawa
couplings real, the constraint is $Y_{e\tau} Y_{\mu\tau} \lesssim 10^{-6}$.
This is almost, but not quite, competitive
with the bound following from $\tau \to e\gamma$ and $\tau \to \mu \gamma$
decays, see Table \ref{tab:leptons}.

\subsection{LEP constraints}

The Large Electron--Positron collider (LEP) is indirectly sensitive to the
flavor violating Yukawa couplings $Y_{\ell e}$ and $Y_{e\ell}$ (with $\ell =
\mu, \tau$) through the process $e^+ e^- \to \ell^+ \ell^-$, mediated by a
$t$-channel Higgs.  The relevant observables here are the total cross sections
$\sigma(e^+ e^- \to \ell^+ \ell^-)$ and the forward--backward asymmetry of the
final state leptons, both of which were measured as a function of the
center-of-mass energy $\sqrt{s}$ with uncertainties of order several per
cent~\cite{Alcaraz:2006mx}.  However, since the new physics contribution to
$\sigma(e^+ e^- \to \ell^+ \ell^-)$
is proportional to four powers of the off-diagonal Yukawa couplings, LEP
limits cannot compete with constraints from flavor-violating decays like $\tau
\to \mu\gamma$ and $\mu \to e\gamma$ and with the LHC constraints we 
 derive in Section~\ref{sec:lhc}.  While a full derivation of LEP limits,
including a careful treatment of the interference between Standard Model and
non-standard contributions as well as a fit to the data points given
in~\cite{Alcaraz:2006mx} is beyond the scope of this work, we have estimated
that flavor-violating couplings $\sqrt{|Y_{\ell e}|^2 + |Y_{e\ell}|^2} \lesssim
\text{few} \times 10^{-1}$ are excluded by LEP.

\subsection{Allowed branching ratios for lepton flavor violating Higgs decays}

\begin{figure}
  \begin{center}
    \vspace{-3mm}
    \includegraphics[width=0.45\textwidth]{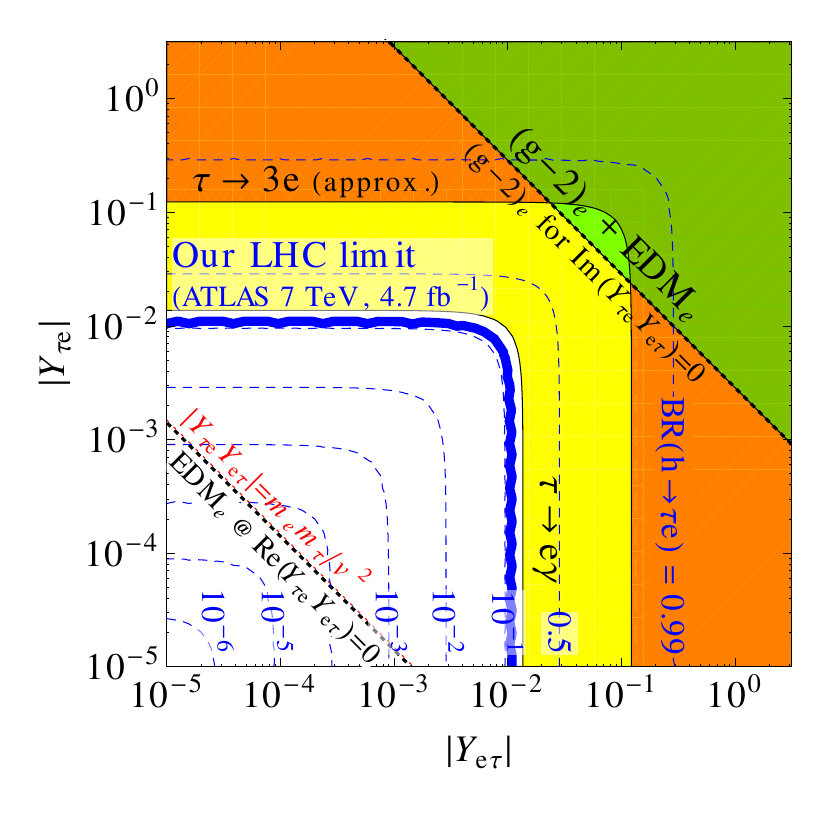} \quad
    \includegraphics[width=0.45\textwidth]{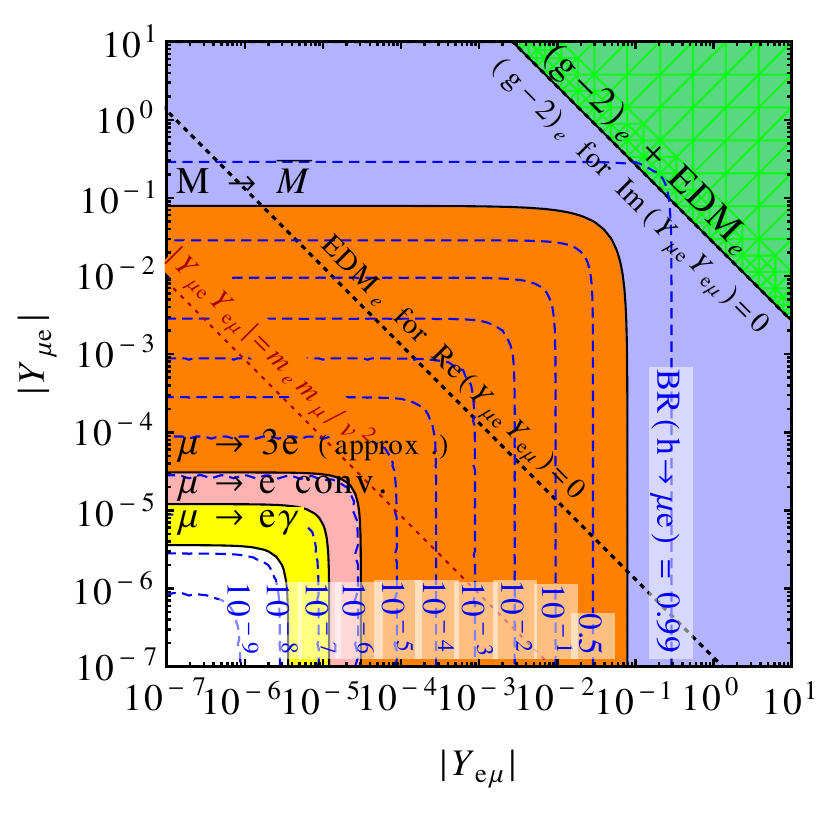}\\[-0.5cm]
    \includegraphics[width=0.45\textwidth]{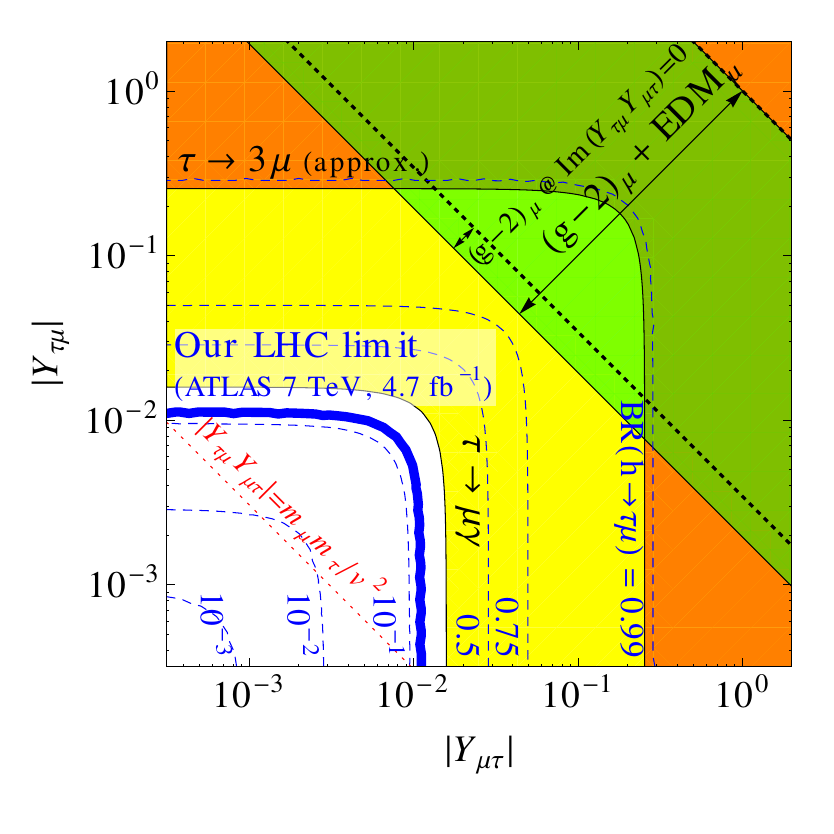}
    \vspace{-13mm}
  \end{center}
  \caption{Constraints on the flavor violating Yukawa couplings $|Y_{e\tau}|$,
  $|Y_{\tau e}|$ (upper left panel), $|Y_{e\mu}|$, $|Y_{\mu e}|$ (upper right
  panel) and $|Y_{\mu\tau}|$, $|Y_{\tau \mu}|$ (lower panel) of a 125~GeV Higgs
  boson. The diagonal Yukawa couplings are approximated by their SM values.
  Thin blue dashed lines are contours of constant BR for $h \to \tau e$, $h \to
  \mu e$ and $h \to \tau\mu$, respectively, whereas thick blue lines are the
  LHC limits derived in Sec.~\ref{sec:lhc-current}.  (These limits could be
  greatly improved with dedicated searches on existing LHC data, see
  Sec.~\ref{sec:lhc-dedicated}.)  Shaded regions show the constraints discussed
  in Sec.~\ref{sec:leptons} as indicated in the plots.  Note that $g-2$ [EDM]
  searches (diagonal black dotted lines) are only sensitive to parameter
  combinations of the form $\Re(Y_{\alpha\beta} Y_{\beta\alpha})$
  [$\Im(Y_{\alpha\beta} Y_{\beta\alpha})$]. We also show limits from a
  combination of $g-2$ and EDM searches with marginalization over the complex
  phases of the Yukawa couplings (green shaded regions). Note that $(g-2)_\mu$
  provides upper and lower limits (as indicated by the double-sided arrows in
  the lower panel) if the discrepancy between the measurement and the SM
  prediction~\cite{Nakamura:2010zzi,Bennett:2006fi} is taken into account.  The
  thin red dotted lines show rough naturalness limits $Y_{ij} Y_{ji}\lesssim
  m_im_j/v^2$ (see Sec.~\ref{sec:framework}).}
  \label{fig:etau-constraints}
\end{figure}

In Fig. \ref{fig:etau-constraints} we collect the above constraints on the
values of   $|Y_{e\tau}|$, $|Y_{\tau e}|$ (upper left panel), $|Y_{e\mu}|$,
$|Y_{\mu e}|$ (upper right panel) and $|Y_{\mu\tau}|$, $|Y_{\tau \mu}|$ (lower
panel) and relate them to the predicted branching ratios for $h\to e\tau$,
$h\to e\tau$ and $h\to \mu\tau$. The latter are given by
\begin{align}
  \BR(h \to \ell^\alpha\ell^\beta) = \frac{\Gamma(h\to \ell^\alpha \ell^\beta)}
    {\Gamma(h\to \ell^\alpha \ell^\beta) + \Gamma_{\rm SM}} \,,
\end{align}
where $\ell^\alpha$, $\ell^\beta$ = $e$, $\mu$, $\tau$, $\ell^\alpha \neq
\ell^\beta$.  The decay width $\Gamma(h\to \ell^\alpha \ell^\beta)$, in turn,
is
\begin{align}
  \Gamma(h \to \ell^\alpha \ell^\beta) = \frac{m_h}{8 \pi} \big(|Y_{\ell^\beta\ell^\alpha}|^2
                                           + |Y_{\ell^\alpha\ell^\beta}|^2 \big) \,,
\end{align}
and the SM Higgs width is $\Gamma_{\rm SM} = 4.1$~MeV for a 125~GeV Higgs
boson~\cite{Dittmaier:2012vm}. In the panels of Fig.~\ref{fig:etau-constraints}
we are assuming that at most one of non-standard decay mode of the Higgs is
significant compared to the SM decay width.

From Fig.~\ref{fig:etau-constraints} we see that given current bounds from
$\tau\to\mu\gamma$ and $\tau\to e \gamma$, branching fractions for $h \to
\tau\mu$ or $h \to \tau e$ in the neighborhood of 10\% are allowed. This is
well within the reach of the LHC as we shall show in Sec.~\ref{sec:lhc}.
The allowed sizes of these two decay widths are comparable to the sizes of
decay widths into nonstandard decay channels (such as the invisible decay
width) that are allowed by global fits \cite{Carmi:2012in}. If there is no
significant negative contribution to Higgs production through gluon fusion, one has
$\BR(h\to \text{invisible})\lesssim 20\%$, while allowing for arbitrarily large
modifications of gluon and photon couplings to the Higgs leads to the constraint $\BR(h\to
{\rm invisible})\lesssim 65\%$ \cite{Carmi:2012in}. These two bounds apply
without change also to $\BR(h\to \tau \mu)$, $\BR(h\to \tau e)$ and $\BR(h\to e
\mu)$.

In contrast to decays involving a $\tau$ lepton, the branching ratio for $h\to
e\mu$ is extremely well constrained by $\mu \to e\gamma$, $\mu \to 3e$ and $\mu\to e$
conversion bounds, and is required to be below $\BR(h\to e\mu)\lesssim2 \times
10^{-8}$, well beyond the reach of the LHC.

\section{Hadronic flavor violating decays of the Higgs}
\label{sec:quarks}

We next consider flavor violating decays of the Higgs to quarks. We first
discuss two-body decays to light quarks, $h\to \bar b d$, $\bar b s$, $\bar{s}
d$, $\bar c u$, and then turn to FV three body decays mediated by an off-shell
top, $h\to \bar t^*c\to W \bar b c$ and $h\to \bar t^*u\to W\bar b u$ as well
as FV top decays to $t\to ch$ and $t\to uh$. Our limits are summarized in
Table~\ref{tab:light-quarks}.

\begin{table}
  \centering
  \parbox{13cm}{
  \begin{ruledtabular}
  \begin{tabular}{@{\qquad}lcc@{\qquad}}
    Technique     & Coupling              &  Constraint              \\ \hline
    \multirow{2}{*}{$D^0$ oscillations~\cite{Bona:2007vi}}
                  & $|Y_{uc}|^2$, $|Y_{cu}|^2$             &  $< 5.0 \times 10^{-9}$ \\
                  & $|Y_{uc} Y_{cu}|$                      &  $< 7.5 \times 10^{-10}$ \\ \hline
    \multirow{2}{*}{$B_d^0$ oscillations~\cite{Bona:2007vi}}
                  & $|Y_{db}|^2$, $|Y_{bd}|^2$             &  $< 2.3 \times 10^{-8}$ \\
                  & $|Y_{db} Y_{bd}|$                      &  $< 3.3 \times 10^{-9}$ \\ \hline
     \multirow{2}{*}{$B_s^0$ oscillations~\cite{Bona:2007vi}}
                  & $|Y_{sb}|^2$, $|Y_{bs}|^2$             &  $< 1.8 \times 10^{-6}$ \\
                  & $|Y_{sb} Y_{bs}|$                      &  $< 2.5 \times 10^{-7}$ \\ \hline
     \multirow{4}{*}{$K^0$ oscillations~\cite{Bona:2007vi}}
                  & $\Re(Y_{ds}^2)$, $\Re(Y_{sd}^2)$       &  $[-5.9 \dots 5.6] \times 10^{-10}$ \\
                  & $\Im(Y_{ds}^2)$, $\Im(Y_{sd}^2)$       &  $[-2.9 \dots 1.6] \times 10^{-12}$ \\
                  & $\Re(Y_{ds}^* Y_{sd})$                 &  $[-5.6 \dots 5.6] \times 10^{-11}$ \\
                  & $\Im(Y_{ds}^* Y_{sd})$                 &  $[-1.4 \dots 2.8] \times 10^{-13}$ \\ \hline
     \multirow{2}{*}{single-top production~\cite{Abazov:2010qk}}
                  & $\sqrt{|Y_{tc}^2| + |Y_{ct}|^2}$       &  $< 3.7$  \\
                  & $\sqrt{|Y_{tu}^2| + |Y_{ut}|^2}$       &  $< 1.6$  \\ \hline
     \multirow{2}{*}{$t\to hj$~\cite{Craig:2012vj}}
                  & $\sqrt{|Y_{tc}^2| + |Y_{ct}|^2}$       &  $< 0.34$ \\
                  & $\sqrt{|Y_{tu}^2| + |Y_{ut}|^2}$       &  $< 0.34$ \\ \hline                  
     \multirow{3}{*}{$D^0$ oscillations~\cite{Bona:2007vi}}
                  & $|Y_{ut} Y_{ct}|$, $|Y_{tu} Y_{tc}|$   &  $< 7.6 \times 10^{-3}$ \\
                  & $|Y_{tu} Y_{ct}|$, $|Y_{ut} Y_{tc}|$   &  $< 2.2 \times 10^{-3}$ \\
                  & $|Y_{ut} Y_{tu} Y_{ct} Y_{tc}|^{1/2}$  &  $< 0.9 \times 10^{-3}$ \\
                  \hline
  neutron EDM~\cite{Beringer:2012PDG,Gorbahn:2014sha} & $|\Im(Y_{ut}Y_{tu})|$    & $<4.3 \times 10^{-7}$\\
                                                      & $|\Im(Y_{ct}Y_{tc})|$    & $<5.0 \times 10^{-4}$\\
  \end{tabular}
  \end{ruledtabular}}
  \caption{Constraints on flavor violating Higgs couplings to quarks.
    We have assumed a Higgs mass $m_h = 125$~GeV, and we have taken the
    diagonal Yukawa couplings at their SM values.}
  \label{tab:light-quarks}
\end{table}

\subsection{Flavor violating Higgs decays into light quarks}
\label{sec:light-quark}

\begin{figure}
  \begin{center}
    \includegraphics[width=0.8\textwidth]{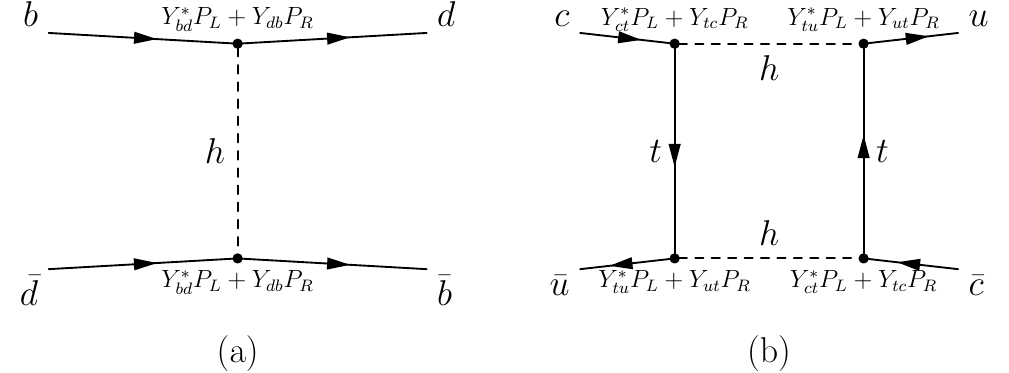}
  \end{center}
  \caption{Two representative diagrams through which flavor violating Higgs Yukawa
    couplings can contribute to neutral meson mixing.}
  \label{fig:meson-osc}
\end{figure}

Flavor violating Higgs couplings to quarks can generate flavor changing neutral
currents (FCNCs) at tree level, see Fig.~\ref{fig:meson-osc}~(a), and are thus
well constrained by the measured $B_{d,s}-\bar B_{d,s}$, $K^0-\bar K^0$ and
$D^0-\bar D^0$ mixing rates.  Integrating out the Higgs generates an effective
weak Hamiltonian, which for $B_d-\bar B_d$ mixing is
\begin{align}
  H_{\rm eff} &= C_2^{db} (\bar b_R d_L)^2 + \tilde C_2^{db}(\bar b_L d_R)^2
               + C_4^{db}(\bar b_L d_R)(\bar b_R d_L) \,.
  \label{eff:weak:Hamiltonian}
\end{align}
Here we use the same notation for the Wilson coefficients as in
\cite{Bona:2007vi} and display only nonzero contributions, which are
\begin{align}
  C_2^{db} &= -\frac{(Y_{db}^*)^2}{2 m_h^2}, \quad
               \tilde C_2^{db}=-\frac{(Y_{bd}^2)^2}{2 m_h^2}, \quad
               C_4^{db}=-\frac{Y_{bd}Y_{db}^*}{m_h^2} \,.
\end{align}
The results for $B_s-\bar B_s$, $K^0-\bar K^0$ and $D^0-\bar D^0$ mixing are
obtained in the same way with the obvious quark flavor replacements.  We can
now translate the bounds on the above Wilson coefficients obtained in
\cite{Bona:2007vi} into constraints on the combinations of flavor violating
Higgs couplings as summarized in Table~\ref{tab:light-quarks}.  We see that all
Yukawa couplings involving only $u$, $d$, $s$, $c$, or $b$ quarks have to be
tiny. The weakest constraints are those in the $b$--$s$ sector, where flavor
violating Yukawa couplings $\lesssim 10^{-3}$ are still allowed.  This would
correspond to $\BR(h \to b s) \sim 2 \times 10^{-3}$, which is still far too
small to be observed at the LHC because of the large QCD backgrounds.

\subsection{Higgs decays through off-shell top and top decays to Higgs}

Among the flavor violating Higgs couplings to quarks, the most promising place
for new physics to hide  are processes involving top quarks, such as the 3-body
decay $h \to (t^* \to W b) q$. Here, $q$ denotes either a charm quark or an up quark.
The corresponding FV Yukawa couplings contribute at one loop to $D-\bar D$ mixing
through diagrams of the form of Fig.~\ref{fig:meson-osc}~(b). The corresponding
Wilson coefficients in the effective Hamiltonian \eqref{eff:weak:Hamiltonian}
are
\begin{align}
  C_1^{uc} &= \frac{1}{4}\frac{1}{16 \pi^2}\frac{S_1^H(x_{tH})}{m_h^2}
              \big(Y_{ct} Y_{ut}^*\big)^2\,, &\tilde
  C_1^{uc} &= \frac{1}{4}\frac{1}{16 \pi^2}\frac{S_1^H(x_{tH})}{m_h^2}
              \big(Y_{tc}^* Y_{tu}\big)^2\,, \\
  C_2^{uc} &=-\frac{1}{4}\frac{1}{16 \pi^2}\frac{S_2^H(x_{tH})}{m_h^2}
              \big(Y_{tc}^* Y_{ut}^*\big)^2\,, &\tilde
  C_2^{uc} &=-\frac{1}{4}\frac{1}{16 \pi^2}\frac{S_2^H(x_{tH})}{m_h^2}
              \big(Y_{ct} Y_{tu}\big)^2\,, \\
  C_4^{uc} &=-\frac{1}{2}\frac{1}{16 \pi^2}\frac{S_2^H(x_{tH})}{m_h^2}
              \big(Y_{ct} Y_{tu}\big)\big(Y_{tc}^* Y_{ut}^*\big)\,, &
  C_5^{uc} &=-\frac{1}{16 \pi^2}\frac{S_1^H(x_{tH})}{m_h^2}
              \big(Y_{ct} Y_{ut}^*\big)\big(Y_{tc}^* Y_{tu}\big) \,,
\end{align}
where
\begin{align}
  S_1^H(x) &= \frac{x^2-1- 2 x \log x}{2 (x-1)^3}\,, \qquad
  S_2^H(x) &= \frac{2 x \big[2- 2x+(1+x) \log x\big]}{(x-1)^3} \,,
\end{align}
and $x_{tH}\equiv m_t^2/m_h^2$. Note that now also the operators $Q_{1,5}^{uc},
\tilde Q_1^{uc}$ (in the notation of \cite{Bona:2007vi}) have non-zero Wilson
coefficients. By requiring that each individual operator is consistent with its
$D-\bar D$ mixing constraint, we derive the limits shown in the last part of
Table~\ref{tab:light-quarks}. The constraints are much weaker than those on FV
Higgs couplings involving only light quarks.

\begin{figure}
  \begin{center}
    \includegraphics[width=0.5\textwidth]{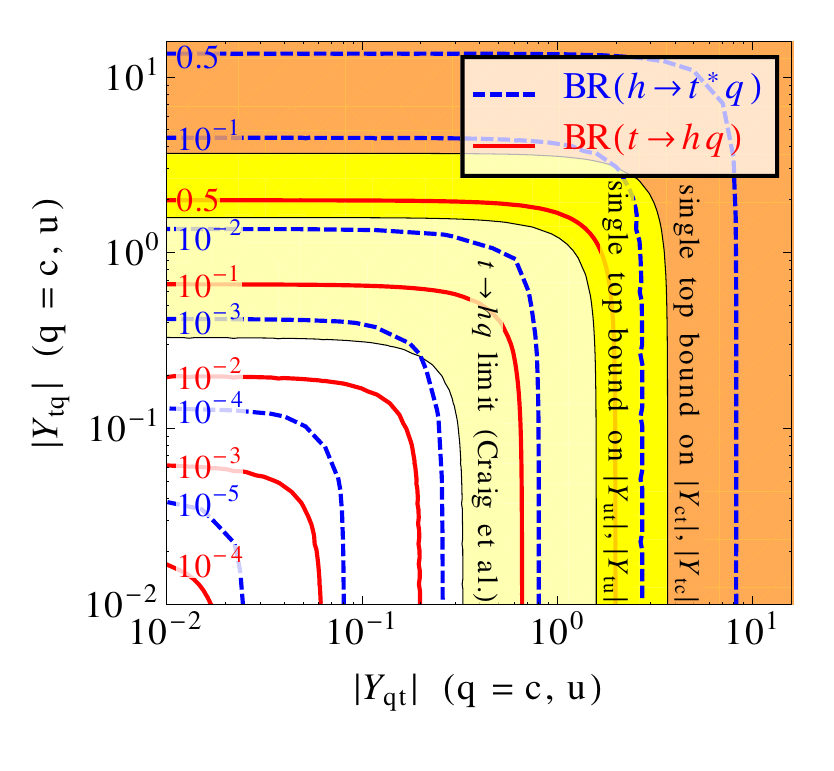}
  \end{center}
  \caption{Predictions for various flavor changing neutral current (FCNC)
  processes mediated by the flavor violating Yukawa couplings $Y_{ct}$,
  $Y_{tc}$ or $Y_{ut}$, $Y_{tu}$ of a 125~GeV Higgs boson.  Where appropriate,
  we have approximated the diagonal Yukawa couplings by their Standard Models
  values.  Blue dashed contours indicate the branching ratio for $h \to t^* q$,
  red solid contours the one for $t \to h q$ (where $q$ denotes a charm or up
  quark). The light yellow region shows a recent limit on $t\to hc$ (or $hu$) from an
  LHC multi-lepton search~\cite{Craig:2012vj}.}
  \label{fig:Yct-constraints}
\end{figure}

Strong constraints on $Y_{qt}$ and $Y_{tq}$ are also obtained from the non-observation
of anomalous single top production. The flavor violating chromomagnetic operators
\begin{align}
  \mathcal{L}_\text{single top} \supset
    \frac{g_s}{m_h} \bar{t} \sigma^{\mu\nu} (\kappa_{tqg,L} P_L + \kappa_{tqg,R} P_R)
    \frac{\lambda^a}{2} q \, G_{\mu\nu}^a \,,
  \label{eq:single-top}
\end{align}
are generated trough loop diagrams similar to Fig.~\ref{fig:tau-mu-gamma}, but
with leptons replaced by quarks and the photon replaced by a gluon.  Here $g_s$
is the strong coupling constant, $\lambda^a$ are the Gell-Mann matrices,
$G_{\mu\nu}^a$ is the gluon field strength tensor, and $\kappa_{tqg,L}$,
$\kappa_{tqg,R}$ are dimensionless effective coupling constants which depend on
$Y_{qt}$ and $Y_{tq}$ according to
\begin{align}
  \kappa_{tqg,L} = \frac{m_t m_h}{8 \pi^2} F(m_t, m_t, 0, q^2=0, Y^\dag) \,,\label{tqg}
\end{align}
with the loop function $F$ given in Eq.~\eqref{eq:tmg-F}.
The analogous expression for $\kappa_{tqg,R}$ is obtained by replacing
$Y_{tq}^*\to Y_{qt}$ and $Y_{tt} \to Y_{tt}^*$ in $F$. Note that in \eqref{tqg}
we have assumed an EFT description with an on-shell gluon. Since $m_h\sim m_t$ this
is only approximate, but we have checked that varying $q^2\in [0,m_t^2]$
changes the bounds on $Y_{tq}$, $Y_{qt}$ only by $\sim 10$\%. We have also
made the approximation $m_q \to 0$, which is obeyed even
much better. Limits on $\kappa_{tqg,L}$, $\kappa_{tqg,R}$
have been derived by the CDF and D\O\ collaborations~\cite{Aaltonen:2008qr,
Abazov:2010qk} and most recently by ATLAS~\cite{Aad:2012gd}. In the notation of
\cite{Aad:2012gd}, we have $|\kappa_{tgf}|/\Lambda \equiv
\sqrt{|\kappa_{tqg,L}|^2+|\kappa_{tqg,R}|^2}/(\sqrt 2 m_h)$. We obtain the
constraints
\begin{align}
  \sqrt{|Y_{tc}^2| + |Y_{ct}|^2} < 3.7\,, \qquad
  \sqrt{|Y_{tu}^2| + |Y_{ut}|^2} < 1.6\,,
\end{align}
We now translate these bounds into constraints on the $h \to (\bar{t}^* \to W
\bar{b}) q$ decay width, which is given by (setting $m_{b,q}=0)$
\begin{align}
  \frac{d^2\Gamma(h \to \bar{t}^* q)}{dm_{12}^2 \, dm_{23}^2} &= \frac{3 g^2 |V_{tb}|^2}{64(2\pi)^3  m_W^2 m_h^3 }
    \, \frac{ 1 }{(m_{23}^2 - m_t^2)^2} \Big[
      m_{12}^2 \big(2 m_W^2 - m_{23}^2 \big) \big(m_t^2 |Y_{qt}|^2 - m_{23}^2 |Y_{tq}|^2 \big)
      \nonumber\\
  &\hspace{3cm}
    + \big(m_h^2 - m_{23}^2 \big) \big(m_{23}^2-m_W^2  \big)
      \big(2 m_W^2 |Y_{tq}|^2 + m_t^2 |Y_{qt}|^2 \big)
    \Big] \,,
  \label{eq:Gamma-htc}
\end{align}
where $V_{tb} \simeq 1$ is a  CKM matrix element.  The branching ratio for
$h\to t^* c$ can be as large as $\mathcal{O}(10^{-3})$, and the one for $h\to t^* u$
can be $\text{few} \times 10^{-4}$ as shown in Fig.~\ref{fig:Yct-constraints}.

If the decay $h \to (t^* \to W b) c$ is non-negligible, so is the related
non-standard top quark decay mode $t \to h c$, the rate for which is given by
(neglecting the charm mass)
\begin{align}
  \Gamma(t \to h c) = \frac{|Y_{ct}|^2 + |Y_{tc}|^2}{32\pi} \,
                      \frac{(m_t^2 - m_h^2)^2}{m_t^3} \,.
  \label{eq:thc}
\end{align}
Branching ratios for $t\to hc$ of several tens of per cent are perfectly viable
and can be searched for, e.g.\ in the multi-lepton or $t\to b\bar b c$
channels. In fact, the strongest hint on Higgs couplings to $tc$ are already
coming from a CMS multi-lepton search which was recast in~\cite{Craig:2012vj}
to search for $t\to hc$, giving a bound of 2.7\% on the branching fraction of a
top into a Higgs and a charm or up quark. This yields a limit of
$\sqrt{|Y_{ti}|^2+|Y_{it}|^2}<0.34$ for $i=u$~or~$c$ (see Fig.
\ref{fig:Yct-constraints}).

We have also calculated the branching ratios for the loop-induced
processes $t \to q \gamma$, $t \to q g$ and $t \to q Z$ ($q = u, c$),
which are in principle sensitive to $|Y_{qt}|$ and $|Y_{tq}|$, but
have found that even for $|Y_{qt}|$, $|Y_{tq}| \sim
\mathcal{O}(1)$ the current experimental bounds are satisfied~\cite{CMSflavor}.

In the above we have assumed that the weak phases of $Y_{ut}$ and $Y_{ut}$ are
negligibly small. Otherwise an unacceptably large contribution to the neutron EDM
is generated at 1-loop level with top and Higgs running in the loop. Eq.~\eqref{eq:dmu}
with the replacements $m_\tau\to m_t$, and $Y_{\mu\tau}Y_{\tau \mu}\to
Y_{ut}Y_{tu}$ gives the $u$-quark EDM $d_u=e \tilde d_u$, from
which one can calculate the neutron EDM $d_n$ \cite{Pospelov:2005pr}. Using the
90\%~CL experimental bound $d_n<0.29 \times 10^{-25} e \, \text{cm}$~\cite{Beringer:2012PDG}
together with the estimate for the relation between the quark and neutron EDMs,
Eq.~(3.62) of \cite{Pospelov:2005pr}, one obtains $|\Im Y_{ut} Y_{tu}| \lesssim 4.4 \times
10^{-8}$. To obtain this limit we have used the full one-loop expression for the
quark EDMs rather than the approximation Eq.~\eqref{eq:dmu}.  For simplicity, we
have neglected the contributions from chromomagnetic operators, which are similar
in magnitude to the terms we keep.
Including chromomagnetic terms and taking into account
renormalization group running as well as using a conservative estimate of
hadronic matrix elements, the authors of~\cite{Gorbahn:2014sha} obtain
$|\Im Y_{ut} Y_{tu}|< 4.3 \times 10^{-7}$ and $|\Im Y_{ct} Y_{tc}|< 5.0 \times 10^{-4}$.

We see that the limit on $|\Im Y_{ut} Y_{tu}|$ is much more stringent than the
bounds on the absolute values
of the same FV Yukawa couplings.  In contrast, our estimates for the bounds from charm running in
the loop, $|\Im Y_{uc} Y_{cu}|< 1.6 \times 10^{-7}$, and from $d$-quark EDMs
generated by the $b$-quark and $s$-quark running in the loop,  $|\Im Y_{db}
Y_{bd}|< 6.4 \times 10^{-8}$ and $|\Im Y_{ds} Y_{sd}|< 1.2 \times 10^{-6}$,
respectively, are less stringent than the bounds from meson mixing, Table
\ref{tab:light-quarks}.

\section{Searching for flavor violating Higgs decays at the LHC}
\label{sec:lhc}

We next discuss possible search strategies for flavor violating Higgs decays at
the LHC, focusing on the  $h \to \tau\mu$ and $h \to \tau e$ decays. As shown
in Fig.~\ref{fig:etau-constraints}, these are among the least constrained of
the couplings discussed in this paper, with a potential to modify the Higgs
branching fractions significantly.  They are sensitive to new particles with
flavor violating couplings or to a secondary mechanisms of electroweak symmetry
breaking such as additional Higgs doublets, and are thus good probes of new
physics.  Furthermore, they are also interesting final states as far as the
potential for searches at the LHC is concerned.

The decay $h \to \tau\mu$ is quite similar to the standard model $h\to\tau\tau$
decay with one of the tau leptons decaying to a muon. This implies that existing SM
Higgs searches, with only small or no modifications at all, can already be used
to place bounds on the flavor violating decay.  We thus first extract limits on
$h \to \tau\mu$ and $h\to \tau e$ decays from an existing $h\to \tau \tau $
search in ATLAS. We then discuss how modifications to the $\tau\tau$ search
can lead to significantly improved sensitivity to flavor violating Higgs decays.

\subsection{Extracting a bound on Higgs decays to $\tau\mu$ and $\tau e$}
\label{sec:lhc-current}

We use the existing ATLAS search for $h\to\tau\tau$ in the fully leptonic
channel~\cite{AtlasHiggsTauTau:2012} to place bounds on the $h\to \tau\mu$ and
$h\to \tau e$ branching fractions. The reason we use fully leptonic events is
that we can simulate the detector response to them more accurately than
for events involving hadronic taus. It should, however, be noted that in the SM
$h \to \tau\tau$ search in ATLAS, semi-hadronic events are about as sensitive as fully
leptonic ones~\cite{AtlasHiggsTauTau:2012}, and in CMS, the semi-hadronic mode provides
even stronger limits~\cite{Chatrchyan:2012vp}. The analysis in~\cite{AtlasHiggsTauTau:2012}
uses the collinear approximation to reconstruct the $\tau\tau$  invariant mass,
i.e.\ it is assumed that the neutrino and the charged lepton emitted in tau decay are collinear.
This approach is less optimized for
$h\to \tau\tau $ than the maximum-likelihood method employed by CMS~\cite{Elagin:2010aw}, but
it is more model independent so that a substantial fraction of
$h \to \tau \mu$ or $h \to \tau e$ decays would pass the cuts.\footnote{In fact, it
may be interesting to apply the collinear approximation more often in resonance
searches. A search for a collinear mass resonance can be sensitive to any
particle which decays to boosted objects whose further decay may introduce
missing energy.} For simplicity we only use the ATLAS cuts optimized for
Higgs production in vector boson fusion (VBF) since this channels provides the
best sensitivity~\cite{AtlasHiggsTauTau:2012}.

To derive limits we have generated 50,000 $pp\to 2j + (h \to \tau\mu)$ Monte
Carlo events using MadGraph~5~v1.4.6~\cite{Alwall:2011uj} for parton level
event generation, Pythia~6.4 for parton showering and hadronization, and
PGS~\cite{PGS} as a fast detector simulation. Combining the ATLAS lepton triggers
and off-line cuts from~\cite{AtlasHiggsTauTau:2012}, we select opposite sign
dilepton events satisfying any of the following requirements: a muon pair with
$p_T > 15$~GeV for the leading muon and $p_T > 10$~GeV for the subleading one,
an electron pair with both $p_T > 15$ GeV, or an electron and a muon with $p_T$
above 15 and 10~GeV, respectively. 
Electrons (muons) are accepted only if their pseudorapidity is
$|\eta| < 2.47 \ (2.5)$.  We require the invariant mass of the lepton pair to
be $30\text{ GeV} < m_{l\bar l} < 100$ GeV for $e\mu$ pairs, or $30\text{ GeV}
< m_{l\bar l}<75$~GeV for same flavor pairs. The missing $p_T$ is required to
be above 20 (40) GeV for $e\mu$ events ($ee$ or $\mu\mu$ events). The azimuthal
separation between the two leptons is required to be $0.5<\Delta\phi_{ll}<2.5$.

Additional cuts are placed with the goal of enriching the event sample in VBF
events: at least two jets with $p_T$ above 40~GeV for the leading jet and above 25~GeV
for the subleading jet are required, with the rapidity difference between the
two leading jets above $|\Delta\eta| > 3$ and the invariant mass $m_{jj} >
350$ GeV. We veto events with an additional jet with $p_T > 25$~GeV and $|\eta|
< 2.4$ in the pseudorapidity region between the two leading jets.

The reconstructed invariant mass is calculated using the collinear
approximation in which all invisible particles are assumed to be collinear with
either of the two leptons. The fractions of the parent $\tau$'s momenta carried
by the charged leptons are denoted by $x_1$ and $x_2$. To be able to compare
with ATLAS data from the $h \to \tau\tau$ search, we compute $x_1$ and $x_2$
assuming \emph{two} neutrinos in the final state, even though $h \to \tau\mu$
yields only one. $x_1$ and $x_2$ are then obtained as the solutions of the
transverse momentum equation $\vec p_{miss,T} = (1-x_1) \vec p_{1,T} + (1-x_2)
\vec p_{2,T}$, where $\vec{p}_{1,2, T}$ are the transverse momenta of the
charged leptons. Following~\cite{AtlasHiggsTauTau:2012}, we require $0.1 <
x_{1,2} < 1$, which removes less than a per cent of $h \to \tau\tau$ events,
but nearly 60\% of our $h \to \tau\mu$ events. Thus, relaxing this cut would
enhance the sensitivity to $h \to \tau\mu$ decays so long as it does not
introduce large backgrounds. Nonetheless, we are still able to use the current
search for $h\to \tau \tau$ to produce an interesting bound on $\BR(h\to \tau
\mu)$.

\begin{figure}
  \begin{center}
    \includegraphics[width=0.49\textwidth]{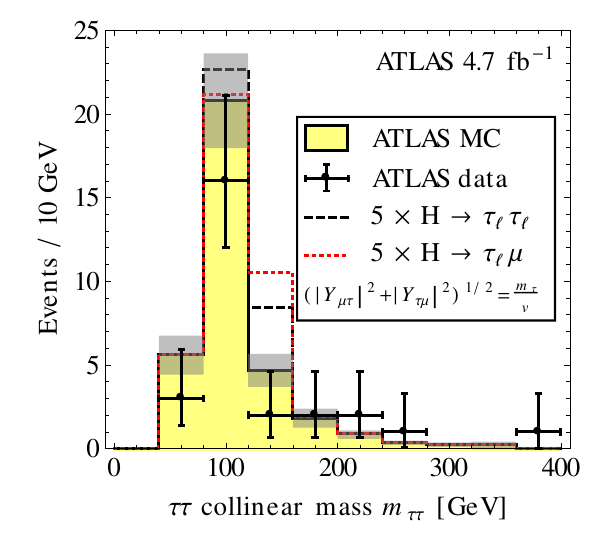}
  \end{center}
  \vspace{-0.7cm}
  \caption{Background rates and $h \to \tau\mu$, $h \to \tau\tau$ signal rates
  in the ATLAS search for fully leptonic $h \to \tau\tau$ decays, optimized for
  Higgs production in vector boson fusion.  The backgrounds expected by
  ATLAS~\cite{AtlasHiggsTauTau:2012} are shown in yellow, with grey bands for
  the systematic uncertainty. Our estimates for the $\tau\mu$ signal at
  $\sqrt{Y_{\tau\mu}^2 + Y_{\mu\tau}^2}=m_\tau/v$ (red) and the SM $h \to
  \tau\tau$ signal (black), which we include for reference, are scaled by a
  factor 5 for illustrative purposes only.}
  \label{fig:ATLAS-search}
\end{figure}

In Fig.~\ref{fig:ATLAS-search} we show the background distribution for the
collinear mass along with the expected shape of a LFV $h\to\tau\mu$ signals
(scaled by a factor five for illustrative purposes only), and we compare to the
observed data. The background expectation is taken
from~\cite{AtlasHiggsTauTau:2012}. The backgrounds and the data in
Fig.~\ref{fig:ATLAS-search} include events for all three combinations of
lepton flavor (even though our $\tau\mu$ signal does not induce $ee$ events)
because only this information is available from ATLAS. For validation purposes,
we have also simulated SM $h \to \tau\tau$ events, and comparing the rate and
shape to Ref.~\cite{AtlasHiggsTauTau:2012} we find agreement to within 20\%.

The $\tau\mu$ signal is predominantly concentrated in the 120--160~GeV bin, so
that the expected and observed limits on the flavor violating Yukawa couplings
can be derived from a simple single-bin analysis.  If we denote the number of
expected background events by $B=4.7$, the number of expected signal events for
a given set of Yukawa couplings by $S$, and the number of observed events by
$O=2$, the expected (observed) one-sided 95\% C.L.\ frequentist limit on $S$ is
defined by the requirement that the probability to observe $\leq B$ ($\leq O$)
events is 5\%. The relevant probability distribution of the data here is a
Poisson distribution with mean $B+S$. We can also include the systematic
uncertainty in the 120--160~GeV bin, which is $\Delta_\mathrm{sys} \simeq \pm
0.99$, in a conservative way by instead using a Poisson distribution with mean
$B + S - \Delta_{\mathrm{sys}}$.  Assuming the Higgs is produced with the
Standard Model rates, this procedure leads to the bound on $\BR(h\to \tau \mu)$
and the analogous bound on $\BR(h\to \tau e)$ shown in
Table~\ref{tab:LHC-tau-mu} (see also Figure~\ref{fig:etau-constraints}).

\begin{table}
  \begin{center}
    \begin{ruledtabular}
    \begin{tabular}{ccccc}
      95\% C.L. limit & $\BR(h\to\tau\mu)$ & $\sqrt{Y_{\tau\mu}^2+Y_{\mu\tau}^2}$ & $\BR(h\to\tau e)$ & $\sqrt{Y_{\tau e}^2+Y_{e\tau}^2}$ \\ \hline
      expected &    28\%     &   0.018     &    27\%     &    0.017  \\
      observed &    13\%     &   0.011     &    13\%     &    0.011
    \end{tabular}
    \end{ruledtabular}
  \end{center}
  \caption{\label{tab:LHC-tau-mu} Expected and observed 95\% C.L.\ limits on
  the $h\to\tau\mu$ and $h\to \tau e$ branching ratios, as well as limits on
  the corresponding Yukawa couplings. The limits are derived by assuming the SM
  Higgs production rates and recasting the search for SM $h \to \tau\tau \to
  2\ell + 2\nu$ decays in the VBF channel from~\cite{AtlasHiggsTauTau:2012}.}
\end{table}

\subsection{Comparison of $h\to\tau\mu$ to $h\to\tau\tau$}
\label{sec:comparison}

We now discuss the experimental differences and similarities between
$h\to\tau\tau$ and $h\to\tau\mu$ decays to determine an optimized search
strategy for the latter. We focus here on $h \to \tau_\text{had}
\tau_\mu$,where $\tau_\mu$ denotes a $\tau$ that decays into a muon and two
neutrinos and $\tau_{\rm had}$ denotes a $\tau$ decaying hadronically. This
channel is actively searched for, both at ATLAS~\cite{AtlasHiggsTauTau:2012}
and at CMS~\cite{Chatrchyan:2012vp}, and is the most sensitive channel in the
CMS $h\to \tau\tau$ search. (In ATLAS, fully leptonic $\tau$ events provide
similar sensitivity to semi-hadronic ones.) It will also be the channel that we
will devise a dedicated search for in the next subsection.

There are a few notable differences between the $h\to \tau_\text{had}\tau_\mu$
and $h\to \tau_\text{had}\,\mu$ decay channels:
\begin{itemize}
  \item {\em Branching Ratios.}
    The branching fraction for $h\to \tau_\text{had}\tau_\mu$ is $2 \times
    \BR(h\to\tau\tau) \times\BR(\tau\to\text{had}) \times \BR(\tau\to\mu)$,
    whereas for $h\to\tau_\text{had}\mu$ it is simply $\BR(h\to\tau\mu) \times
    \BR(\tau\to\text{had})$. For $(Y_{\tau\mu}^2 + Y_{\mu\tau}^2)^{1/2}
    \sim Y_{\tau\tau}$ the signal for
    $h\to \tau_\text{had}\mu$ is  thus a factor of $\sim 2.9$ larger.

  \item {\em Lepton Flavor.}
    The flavor violating decays can lead to different rates for muons and
    electrons in the final state, whereas $\tau\tau$ decays lead to equal $\mu$
    and $e$ rates. Thus, if the various lepton flavor combinations were studied
    separately in the $h \to \tau\tau$ analyses, stronger bounds on flavor
    violating decays could be inferred.

  \item {\em Kinematics and Efficiencies.}
    In $h\to\tau_\text{had}\tau_\mu$ decays the muon carries an average energy
    $\sim m_h/6$, while for  $h\to \tau_{\rm had}\mu$ it carries $\sim m_h/2$.
    Furthermore, in $h\to \tau_\text{had}\,\mu$ events the missing energy is
    roughly aligned with the hadronic $\tau$. As a result the two channels can
    have different efficiencies given the same cuts. For example, in the VBF
    analysis described below (mimicking~\cite{Chatrchyan:2012vp}) the
    efficiency for $h\to \tau_\text{had}\tau_\mu$ is a factor of $\sim 1.8$
    lower than for $h\to \tau_\text{had}\mu$ events, mostly because many of the
    muons in the $h\to \tau_\text{had}\tau_\mu$ sample fall below the $p_T <
    17$~GeV cut.
    
  \item {\em Mass reconstruction.}
    The LHC collaborations use highly optimized procedures for reconstructing
    the $\tau_\text{had} \tau_\mu$ invariant mass. ATLAS uses the Missing Mass
    Calculator (MMC) from~\cite{Elagin:2010aw}, while CMS uses an
    in-house maximum likelihood analysis~\cite{Chatrchyan:2012vp}. These
    procedures use $\vec{p}_{\text{miss},T}$ and the 3-momenta of the muon and
    the $\tau$ jet as input and estimate the neutrino momenta by assuming
    typical $\tau$ decay kinematics. For $h\to \tau\tau$ events, the MMC
    procedure returns an invariant mass with high efficiency ($\sim97\%$) and
    gives a Higgs mass resolution of $\sim 20\%$. If the event is not from
    $h\to \tau\tau$ but instead from $h\to\tau_\text{had}\,\mu$, then i)  the
    efficiency will be significantly lower since the kinematics can be
    completely inconsistent with a $\tau\tau$ event, and ii) the reconstructed
    Higgs mass will be significantly higher as the MMC will assume that the
    hard muon is accompanied by two roughly collinear and hard neutrinos. This
    illustrates that a mass reconstruction procedure designed for the specific
    final state under consideration is mandatory to obtain the best possible
    sensitivity.

  \item {\em Backgrounds.}
    The backgrounds for $h\to \tau_\text{had}\tau_\mu$ and $h\to
    \tau_\text{had}\,\mu$ events are similar, but because of the different
    invariant mass reconstruction techniques, the reconstructed background
    spectra will typically be harder for a $h \to \tau_\text{had}\tau_\mu$
    analysis, which assumes three neutrinos in the final state, than for a $h
    \to \tau_\text{had} \mu$ analysis which assumes only one. This implies, for
    instance, that the peak from the $Z\to \tau\tau$ background will appear at
    a $\tau_\text{had} \tau_\mu$ invariant mass around 90~GeV in a search for
    $h \to \tau_\text{had}\tau_\mu$, but well below (and thus further away from
    the signal peak) in a dedicated $h \to \tau_\text{had} \mu$ analysis.
\end{itemize}
These considerations show that the LHC is potentially more sensitive to flavor
violating $h \to \tau_\text{had} \mu$ decays than to the SM $h \to \tau\tau$
channel.
We now discuss a possible strategy for a tailored $h \to \tau_\text{had} \mu$
analysis.

\subsection{A dedicated $h\to\tau\mu$ analysis}
\label{sec:lhc-dedicated}

We now investigate the potential of a dedicated $h \to \tau_\text{had} \mu$
analysis which follows closely the CMS search for $h \to \tau_\text{had}
\tau_\mu$~\cite{Chatrchyan:2012vp}.\footnote{For semileptonically decaying
$\tau$, search strategies similar to the ones investigated in
Ref.~\cite{Davidson:2012wn} for flavor violating $Z\to \tau\mu$ decays could
be used.}  The most important difference to that
analysis will be a different algorithm for reconstructing the $\tau\mu$
invariant mass. In particular, since the $\tau_\text{had} \mu$ final state
contains only one neutrino (from the hadronic $\tau$), this mass reconstruction
can always be done exactly (i.e.\ the neutrino momentum can be determined) up
to a two-fold ambiguity.

An important background for $h \to \tau\mu$ is $Z + \text{jets}$, where
either the $Z$ decays into $\tau^+ \tau^-$ and one of the $\tau$'s decays
further into a muon, or the $Z$ decays into $\mu^+ \mu^-$ and one of the jets
fakes a $\tau$. Another important background is $W + \text{jets}$, followed by
$W \to \mu \nu_\mu$ and a jet faking a $\tau$.
We neglect the small $t \bar{t}$ background, where a final state $\tau$ can
come from a $W$ decay or be faked
by a jet, and a muon can originate from a $W$ decay or from a leptonic $\tau$
decay. We also do not consider backgrounds from QCD multijet production because
making reliable predictions for these events requires full detector simulations. Based on
the CMS $h \to \tau_\mu \tau_{\rm had}$ search~\cite{Chatrchyan:2012vp} we
expect them to be about as large as the $W + \text{jets}$ background
in the invariant mass region around 125~GeV.

To simulate the parton-level signal and background events, we use
MadGraph~5~v1.4.6 \cite{Alwall:2011uj}, with an extended version of the Higgs
Effective Theory model to include flavor-violating Higgs interactions. We use
Pythia~6.4 for parton showering and hadronization and
Delphes~2.0.2~\cite{Ovyn:2009tx} as a fast detector simulation.  We have
compared the $\tau$ detection efficiency as well as the fake rate from QCD jets
in Delphes~2.0.2 to the corresponding performance indicators of several CMS
$\tau$ tagging algorithm. With a tagging efficiency of $\sim 0.2$ and a fake
rate between 0.2\% at low $p_T$ and 1\% at $p_T \gtrsim 60$~GeV, Delphes
somewhat underestimates the performance of the CMS $\tau$ tagging
algorithms~\cite{CMS-PAS-TAU-11-001}. (We have also studied $\tau$ tagging in
PGS~\cite{PGS}, but found it to be even farther away from what CMS can
achieve.) To compensate for the imperfections in our treatment of
$\tau$-tagging, and for other inaccuracies in our simulations, we normalize our
background distributions to the expected event numbers
from~\cite{Chatrchyan:2012vp}, Table~2. We also normalize the $h \to \tau\mu$ signal
using the same scaling factor as for the SM $h \to \tau_{\rm had} \tau_\mu$
events.

In the analysis we require exactly one muon with $p_{T} > 17$~GeV and $|\eta| <
2.1$ in the final state and exactly one jet tagged as a hadronic $\tau$ decay
with $p_{T} > 20$~GeV and $|\eta| < 2.3$. The muon and the $\tau$ are required
to have opposite charge.  In \cite{Chatrchyan:2012vp}, it was found that the
best signal-to-background ratio is achieved in the events where the Higgs boson
was produced through vector boson fusion (VBF), and we confirm this in our own
simulations. To enrich the data sample in VBF events, we consider only events
with a pair of jets $j_1$, $j_2$ satisfying $|\Delta\eta| > 4.0$, $\eta_1
\eta_2 < 0$, $m_{jj} > 400$~GeV and no other jets with $p_T > 30$~GeV in the
pseudorapidity region between the $j_1$ and $j_2$. Here, $\Delta\eta = \eta_1 -
\eta_2$ is the pseudorapidity difference between the two jets and $m_{jj}$ is
the invariant mass of the jet pair.  Non-$\tau$ jets are included in the
analysis so long as their $p_T$ is above 30~GeV and their pseudorapidity is
$|\eta|<4.7$.  In the CMS analysis~\cite{Chatrchyan:2012vp}, the transverse
mass of the muon and the missing energy is restricted to be below 40~GeV in
order to suppress the $W+\text{jets}$ background. This works because in
$W+\text{jets}$ events, the muon and the neutrino from $W \to \mu \nu_\mu$ tend
to be more back-to-back than in $h \to \tau_\mu \tau_{\rm had}$, where both
$\tau$'s contribute to the missing energy. In $h \to  \tau_{\rm had} \mu$,
however, the muon and the missing energy also tend to be back-to-back, so that
the $m_T(\mu, \vec{p}_{\text{miss},T})$ cut also removes a large fraction of signal
events.

\begin{figure}
  \begin{center}
       \includegraphics[width=0.49\textwidth]{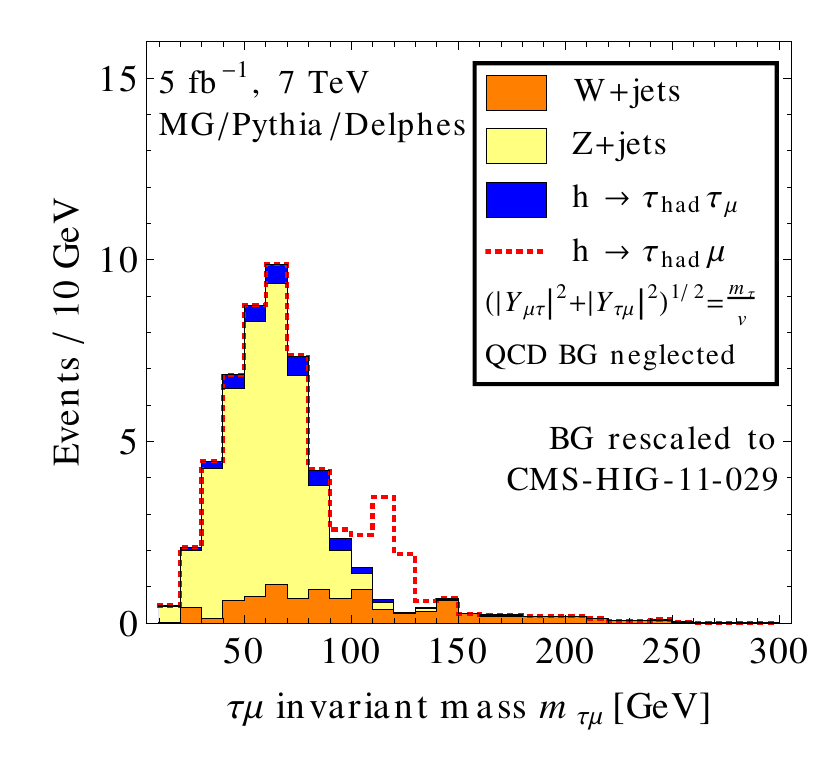}
       \includegraphics[width=0.49\textwidth]{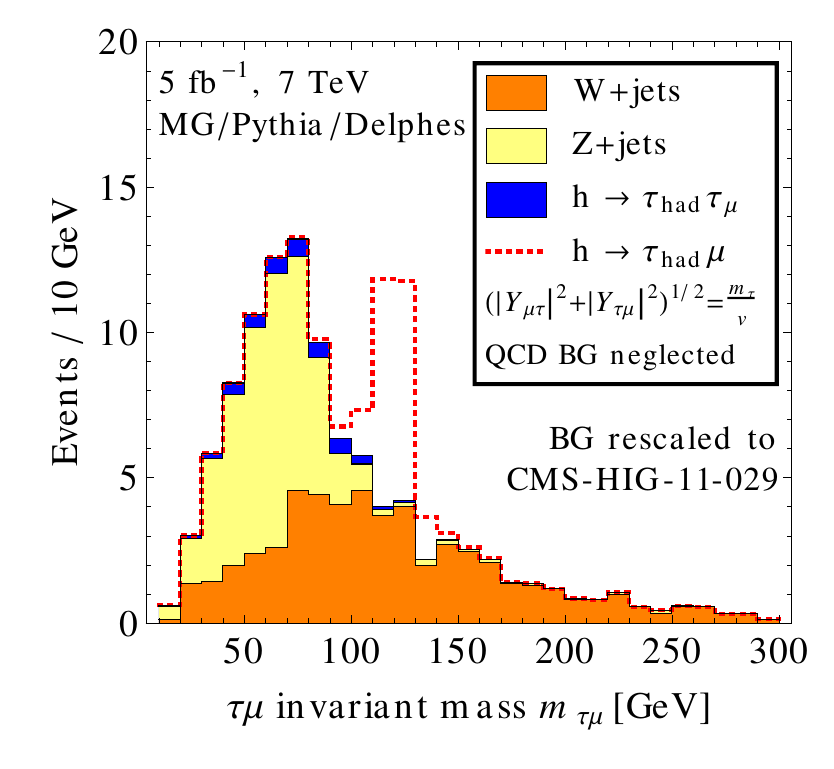}
  \end{center}
  \vspace{-0.7cm}
  \caption{Signal and background rates for $h \to \tau\mu$ events in a CMS-like
  search (see text) as a function of the reconstructed $\mu$--$\tau$ invariant
  mass $m_{\tau\mu}$ for a vector boson fusion-enriched event sample. In the
  left panel the transverse mass cut $m_T(\mu, \vec{p}_{\text{miss},T}) <
  40$~GeV is included, while in the right panel it is omitted.  The QCD
  multijet background and the small $t \bar{t}$ background, are not included.
  The value chosen for
  $\sqrt{Y_{\tau\mu}^2 +Y_{\mu\tau}^2}$ is well within the region allowed by
  other searches for flavor violation in the $\mu$--$\tau$ system (see
  Sec.~\ref{sec:leptons}).}
  \label{fig:htaumu-events}
\end{figure}

\begin{figure}
  \begin{center}
    \includegraphics[width=0.5\textwidth]{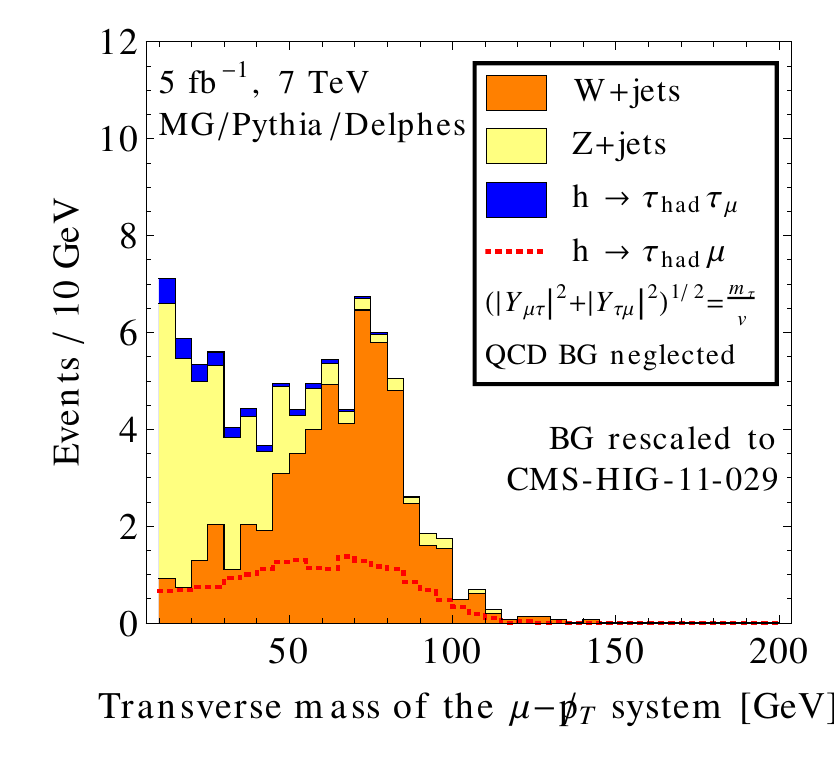}
  \end{center}
  \vspace{-0.7cm}
  \caption{The transverse mass distribution of the muon--missing energy system
    for the backgrounds and for the $\tau_\text{had}\mu$ signal.}
  \label{fig:transverse-mass}
\end{figure}

In light of this we show in Fig.~\ref{fig:htaumu-events} the expected signal
and background rates for $h \to \tau\mu$ as a function of the $\mu$--$\tau$
invariant mass $m_{\tau\mu}$ both with and without the transverse mass cut.  In
computing $m_{\tau\mu}$ for each event, we have used energy and momentum
conservation to compute the $z$-component of the neutrino momentum $p_{\nu,z}$.
There are two solutions to these equations, and we arbitrarily pick the smaller
of the two. (We have checked that choosing the larger value for $p_{\nu,z}$
yields a very similar plot.  This is related to the fact that $m_\tau \ll m_h$,
so that the $\tau$'s decay products are almost collinear.) As shown in the
right panel of Fig.~\ref{fig:htaumu-events}, dropping the transverse mass cut
increases the $W$ plus jets background, but has the benefit of retaining more
signal. The transverse mass distributions for signal and background is shown in
Fig.~\ref{fig:transverse-mass}. Assuming the $W$  plus jets and QCD backgrounds
can be controlled reasonably, relaxing this cut may be worthwhile.

In summary, Fig.~\ref{fig:htaumu-events} shows that for flavor violating Yukawa
couplings well allowed by low energy precision measurements, a spectacular
signal can be expected in a dedicated search at the LHC. Such a search would cut
deeply into the allowed parameter space of the flavor violation Higgs to
$\tau\mu$ couplings.

\section{Conclusions}
\label{sec:conclusions}

The LHC experiments have recently discovered a Higgs-like resonance with a mass
around 125~GeV. In this paper we have examined the constraints on potential
flavor violating couplings of this resonance, assuming it is indeed a scalar boson. In
deriving the constraints we have assumed that that flavor changing neutral
currents are dominated by the Higgs contributions, which may be thought of as
a ``simplified model'' approach to flavor violation in light of the Higgs discovery.
(In a complete model, cancellations between Higgs-induced flavor violation and
flavor violation induced by other new physics is possible, but we do not pursue
this possibility here.)

We have refined the indirect constraints on the flavor violating Yukawa couplings
$Y_{ij}$ using results from rare decay searches, magnetic and electric dipole
moment measurement, and the LHC. All constraints are summarized in
Tables~\ref{tab:leptons} and~\ref{tab:light-quarks} and in
Figs.~\ref{fig:etau-constraints} and \ref{fig:Yct-constraints}. We have
compared the bounds to the loose    naturalness criterion that the off-diagonal
Yukawa couplings are not much bigger then the geometric mean of diagonal terms,
$Y_{ij}\lesssim \sqrt{Y_{ii} Y_{jj}}$,  and we have discussed to what extent
the LHC can probe flavor violating decays of the form $h\to \bar f_i f_j$ (or
in the case of flavor violating top--Higgs couplings the decay $t\to h f_i$).

We draw the following conclusions:

\paragraph*{Natural Flavor Violation.}
The existing constraints involving only the first two generations of fermions,
quarks or leptons, are strong enough that natural FV is already being probed by
meson oscillations, $\mu\to e$ conversion and $\mu \to e \gamma$. This conclusion also holds for FV
couplings to $b$ quarks. In contrast, the FV couplings involving $\tau$ leptons
or top quarks are allowed to have natural size (unless there is a large
hierarchy between $Y_{ij}$ and $Y_{ji}$). This means that they are potentially
observable, either at the LHC or in future low energy experiments.

\paragraph*{Opportunities for the LHC.}
The LHC has an opportunity to probe a large part of the allowed parameter space
for $h \to \tau\mu$ and $h\to \tau e$ couplings.  An LHC search in these
channels would be very similar to the existing searches for $h \to \tau\tau$,
and recasting the latter already gives the best bounds on the flavor violating
Yukawa couplings $Y_{\tau e}$, $Y_{e\tau}$, $Y_{\tau\mu}$, and $Y_{\mu\tau}$
already now. A dedicated LHC search could improve the limits significantly. The
reason why Higgs decays are very constraining is that the SM width of a 125~GeV
Higgs boson is very small, $\Gamma_h \simeq 4$~MeV~\cite{Dittmaier:2012vm},  so
that the flavor violating couplings of the Higgs can have a significant effect.
Another illustration of the LHC's discovery potential for flavor violating
Higgs couplings is that even the global fits of potential deviations in the
dominant SM Higgs decay modes, $h\to WW, ZZ, b\bar b, \tau \tau, \gamma\gamma$,
already give meaningful bounds on the FV Higgs decays. The results from these
global fits are usually presented as bounds on the invisible decay width of the
Higgs, but these bounds applies equally well to the sum of all the modes that
have not been included in the fits. The constraint $\BR(h\to {\rm
invisible})\lesssim {\mathcal O}(70\%)$ (at 95 \% CL, with modest theory
assumptions \cite{Carmi:2012in,Espinosa:2012vu,Djouadi:2012zc,Giardino:2012dp})
is comparable to the constraints on $\BR(h\to \tau \mu)$ and $\BR(h\to \tau e)$
from precision searches of FCNCs in the lepton sector.

Finally, flavor violating Higgs couplings involving the top quark are poorly
constrained, and in fact, the LHC already provides the strongest limits on such
couplings, see Fig. \ref{fig:Yct-constraints} and Section~\ref{sec:quarks}.

\section*{Acknowledgments}

We thank Edward Boos, Alejandro Celis, Andreas Crivellin, Zackaria Chacko,
Ricky Fok, Graham Kribs, Uli Haisch, Ethan Neil, Takemichi Okui, Andr\'e
Sch\"oning and Ze'ev Surujon for valuable discussions, and LOT Polish Airlines
for free internet access during a crucial phase of this project. This material
is based upon work supported in part by the National Science Foundation under
Grant No.  PHYS-1066293 and the hospitality of the Aspen Center for Physics. JZ
was supported in part by the U.S. National Science Foundation under CAREER
Grant PHY1151392. Fermilab is operated by Fermi Research Alliance under
contract DE-AC02-07CH11359 with the United States Department of Energy.

\appendix
\section{Further details on leptonic FCNCs}
\label{Appendix:FCNC}

In this appendix we collect detailed expressions for the FCNC processes $\tau
\to \mu \gamma$ and $\mu \to e$ conversion in nuclei.

\subsection{One loop expressions for $\tau \to \mu \gamma$,  $\tau \to e \gamma$, $\mu\to e\gamma$}

The $\tau \to \mu\gamma$ effective Lagrangian is given in Eq.
\eqref{eq:O-tau-mu-gamma}.  The Wilson coefficients $c_{L,R}$ are given by
\begin{align}
  c_L &= F(m_\tau, m_\tau, m_\mu, m_h, 0, Y) + F(m_\tau, m_\mu, m_\mu, m_h, 0, Y) \,,
  \label{eq:tmg-cL} \\
  c_R &= F(m_\tau, m_\tau, m_\mu, m_h, 0, Y^\dag) + F(m_\tau, m_\mu, m_\mu, m_h, 0, Y^\dag) \,,
  \label{eq:tmg-cR}
\end{align}
with the loop functions
\begin{align}
\begin{split}
  F(m_i, m_f, m_j, m_h, q^2, Y)
    &= \frac{1}{4 m_i} \int_0^1 \! dx\,dy\,dz\,\delta(1-x-y-z) \\
    &\hspace{2.2cm}
        \frac{  x z \, m_j Y_{jf} Y_{if}^*
              + y z \, m_i Y_{fj}^* Y_{fi}
              + (x + y) m_f Y_{fj}^* Y_{if}^*}
              {z m_h^2 - x z \, m_j^2 - y z \, m_i^2 + (x+y) m_f^2 - x y q^2} \,.
  \label{eq:tmg-F}
  \end{split}
\end{align}
Here $m_\mu$, $m_\tau$ and $m_h$ are the muon, tau and Higgs masses,
respectively, $q$ is the 4-momentum of the photon, and $Y$ is the Yukawa
coupling matrix.  Note that $c_L$ and $c_R$ differ only by the replacement
$Y_{ij} \leftrightarrow Y_{ji}^*$. The first terms in Eqs.~\eqref{eq:tmg-cL}
and \eqref{eq:tmg-cR} arise from the first diagram in
Fig.~\ref{fig:tau-mu-gamma} (with a $\tau$ propagating in the loop), whereas
the second terms arise from the second diagram (with a $\mu$ in the loop).
Expanding in powers of $m_\mu/m_\tau$ and $m_\tau/m_h$ and keeping only the
leading terms (so that only the first terms in \eqref{eq:tmg-cL},
\eqref{eq:tmg-cR} contribute), the above expressions simplify to
\eqref{eq:tmg-CR-simplified} if the diagonal Yukawa couplings are real.  The simplified
expressions for $\tau\to e\gamma$ and $\mu \to e\gamma$ (with a muon running in
the loop) are obtained from \eqref{eq:tmg-CR-simplified} with trivial
modifications, while the simplified expression for $\mu\to e\gamma$ with a
$\tau$ running in the loop is given in Eq. \eqref{eq:tmg-CL-simplifiedmue}.

\subsection{Two loop expressions for $\tau \to \mu\gamma$, $\tau \to e\gamma$
            and $\mu\to e\gamma$ \label{App:mue-2loop}}

\begin{figure}
  \begin{center}
    \includegraphics[width=0.75\textwidth]{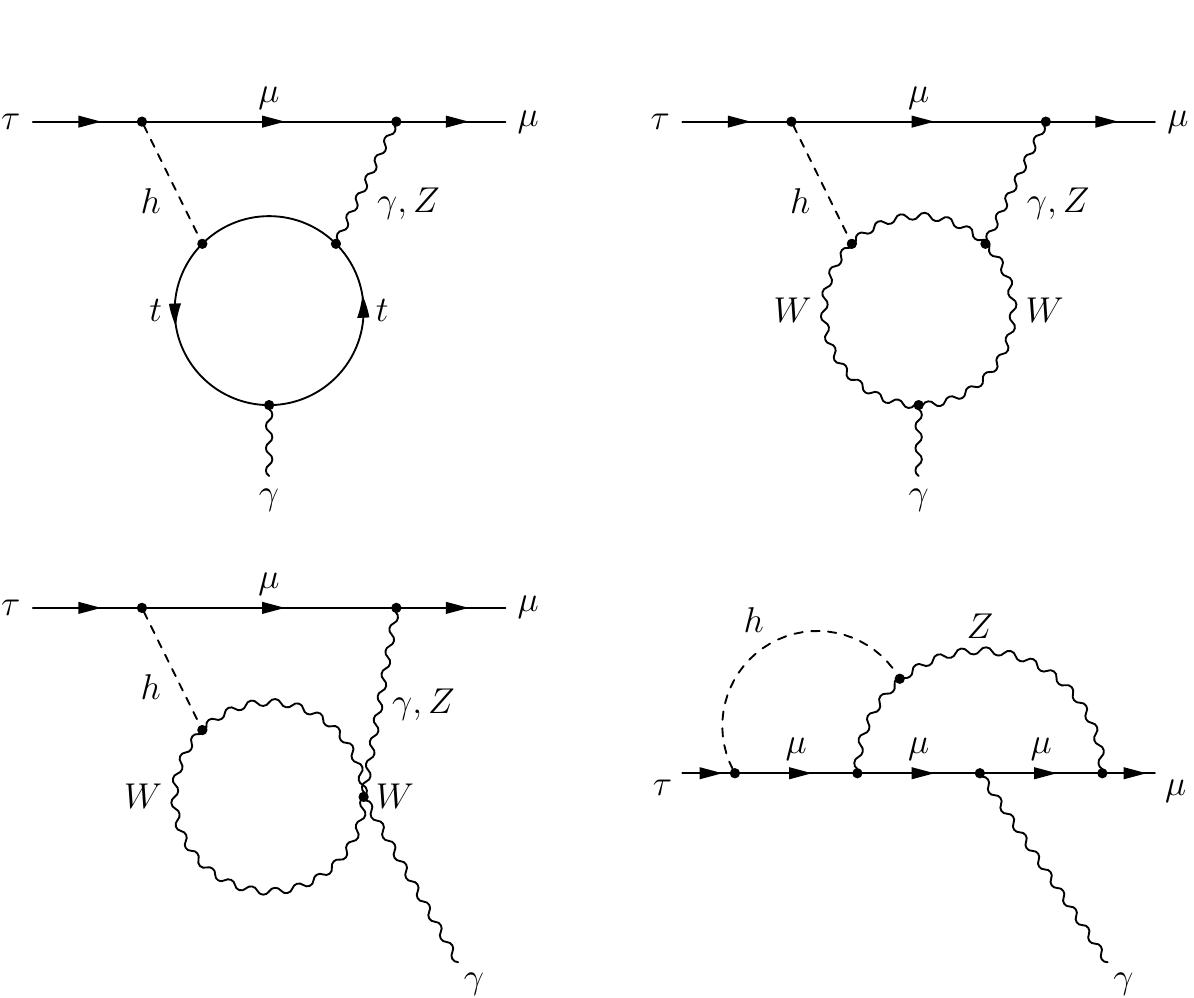}
  \end{center}
  \vspace{-0.7cm}
  \caption{ The two loop diagrams contributing to $\tau \to \mu\gamma$.}
  \label{fig:2loop}
\end{figure}

At two loops there are numerically important diagrams with top or $W$ running
in the loop, attached to the   Higgs. Here we translate the results of
\cite{Chang:1993kw} into our notation and adapt them to the case of   $\tau \to
\mu\gamma$. The diagrams with top and photon in the loops (see
Fig.~\ref{fig:2loop} top left) contributes as
\begin{align}
  \Delta c_L^{t\gamma} &= -6 \kappa Q_t^2 \frac{ v}{m_t}  Y_{\tau \mu}^*
                          \big[\Re (Y_{tt}) f(z_{th})-i \Im(Y_{tt}) g(z_{th})\big]\,,
\end{align}
while the $W$-photon 2-loop contribution is
\begin{align}
  \begin{split}
    \Delta c_L^{W\gamma} &= {\kappa} Y_{\tau \mu}^*\Big[ 3 f(z_{Wh})+5 g(z_{Wh}) +
                            \tfrac{3}{4}g(z_{Wh}) + \tfrac{3}{4}h(z_{Wh}) + \frac{f(z_{Wh})
                            - g(z_{wh})}{2z_{Wh}}\Big] \,.
    \end{split}
\end{align}
Here we have already added the contributions from the would-be Goldstone bosons
that get eaten by the $W$.  The   contributions to $\Delta c_R^i$ are obtained
from the above by replacing $Y_{\tau \mu }^*\to Y_{\mu\tau}$ and $Y_{tt}\to
Y_{tt}^*$.The loop functions are
\begin{align}
  f(z) &= \frac{1}{2} z \int_0^1 dx \frac{1-2x(1-x)}{x(1-x)-z}\log\frac{x(1-x)}{z} \,,\\
  g(z) &= \frac{1}{2} z \int_0^1 dx \frac{1}{x(1-x)-z}\log\frac{x(1-x)}{z} \,,\\
  h(z) &= z^2\frac{\partial}{\partial z}\Big(\frac{g(z)}{z}\Big)
        = \frac{z}{2}\int_0^1 \frac{dx}{z-x(1-x)} \Big[1 + \frac{z}{z-x(1-x)}\log\frac{x(1-x)}{z}\Big]\,,
\end{align}
the arguments are $z_{th}=m_t^2/m_H^2$, $z_{Wh}=m_W^2/m_H^2$, while the prefactor is
\begin{align}
  \kappa &= \frac{\alpha}{16\pi}\frac{g^2}{m_W^2} \frac{v}{m_\tau}
          = \frac{\alpha}{2\sqrt2\pi}G_F \frac{v}{m_\tau}.
\end{align}

The contributions from the 2-loop diagrams with an internal $Z$ are smaller as
they are suppressed by $1 - 4 s_W^2\simeq0.08$. They are
\begin{align}
  \begin{split}
    \Delta c_L^{t Z} &= -6 \kappa Q_t \frac{(1-4 s_W^2)(1-4 Q_t s_W^2)}{16 s_W^2 c_W^2}
                         \frac{v}{m_t} Y_{\tau \mu}^*\times \\
                     &\qquad\qquad\times
                        \big[ \Re(Y_{tt}) \tilde f(z_{th},z_{tZ})
                          - i \Im(Y_{tt})\tilde g(z_{th},z_{tZ})\big] \,,
  \end{split} \\
  \begin{split}
    \Delta c_L^{WZ} &= {\kappa} \frac{1-4 s_W^2}{4 s_W^2} Y_{\tau \mu}^*
                       \Big\{\tfrac12(5-t_W^2) \tilde f(z_{th}, z_{WZ})
                     + \tfrac12(7-3t_W^2)\tilde g(z_{th},z_{WZ}) \\
                    &\qquad
                     + \tfrac34 g(z_{th})+\tfrac34 h(z_{th})
                     + \tfrac{1}{4z_{th}}(1-t_W^2) \big[\tilde f(z_{th},z_{WZ})
                     - \tilde  g(z_{th},z_{WZ})\big]\Big\} \,,
\end{split}
\end{align}
with $s_W\equiv \sin\theta_W$, $c_W\equiv \cos \theta_W$, $t_W\equiv \tan
\theta_W$, $z_{tz}\equiv m_t^2/     m_Z^2$, $z_{WZ}\equiv m_W^2/m_Z^2$ and the
loop functions
\begin{align}
  \tilde f(x,y) &= \frac{y f(x)}{y-x} + \frac{x f(y)}{x-y} \,, \qquad\qquad
  \tilde g(x,y)  = \frac{y g(x)}{y-x} + \frac{x g(y)}{x-y}.
\end{align}
The $\Delta c_R^i$ are obtained by replacing $Y_{\mu e}^*\to Y_{e\mu}$ and
$Y_{tt}\to Y_{tt}^*$ in the above expressions. In addition there are also
contributions called ``set C'' in~\cite{Chang:1993kw}. An example for
one of these diagrams is the last diagram in Fig.~\ref{fig:2loop}. Since the
expressions for these diagrams are long we do not write them out explicitly.
The ``set C'' contribution to $\Delta c_L$ is obtained from \cite{Chang:1993kw}
by multiplying their Eq.~(20) by $-\kappa$ and replacing $\sum_a \cos\varphi_a
\Delta_{e\mu}^a\to Y_{\tau\mu}^*$.

The $\tau \to e\gamma$ expressions are obtained by replacing $Y_{\tau \mu}\to
Y_{\tau e}$ and $Y_{\mu \tau}\to Y_{e \tau}$ in the above expressions, while
for $\mu \to e \gamma$ the replacements are $Y_{\tau \mu}\to Y_{\mu e}$,
$Y_{\mu \tau}\to Y_{e \mu}$ and $m_\tau \to m_\mu$.

\subsection{Details on $\mu\to e$ conversion bounds}
\label{App:mutoe}

The most general effective Lagrangian for $\mu \to e$ conversion in nuclei
is~\cite{Kitano:2002mt}
\begin{align}
  \begin{split}\label{eq:mue:eff}
    {\cal L} &= c_L \frac{e}{8\pi^2} m_\mu(\bar e \sigma^{\alpha \beta}P_L\mu) F_{\alpha \beta}
              - \frac{1}{2} \sum_q \Big[g_{LS}^q (\bar eP_R \mu) (\bar q q)
              + g_{LP}^q (\bar e P_R \mu) (\bar q \gamma_5 q)  \\
             &+ g_{LV}^q (\bar e\gamma^\mu P_L \mu) (\bar q \gamma_\mu q)
              + g_{LA}^q (\bar e\gamma^\mu P_L \mu) (\bar q \gamma_\mu \gamma_5 q)
              + \frac{1}{2} g_{LT}^q (\bar e \sigma^{\alpha \beta}P_R \mu)
                                     (\bar q \sigma_{\alpha \beta}q)\Big] + L\leftrightarrow R.
  \end{split}
\end{align}
The Wilson coefficients $c_L$ and $c_R$ of the magnetic dipole operator are
the same as the ones introduced for $\mu \to e \gamma$ in  Appendix~\ref{App:mue-2loop},
with the replacements $\tau \to \mu$, $\mu \to e$.  They receive contributions
from one-loop and two-loop diagrams, with the two-loop diagrams being orders of magnitude
larger numerically.

The scalar operators in Eq.~\eqref{eq:mue:eff},
generated by the first diagram in Fig.~\ref{fig:mu-e-conversion}, are given by
\begin{align}
  g_{LS}^q = -\frac{2}{ m_h^2}Y_{e \mu} \Re\big(Y_{qq}\big)\,, \qquad
  g_{RS}^q = -\frac{2}{ m_h^2}Y_{\mu e}^* \Re\big(Y_{qq}\big)\,.
\end{align}

The vector
operators are determined at one-loop by the last two diagrams in Fig.~\ref{fig:mu-e-conversion},
with either a muon or an electron running in the
loop.   Explicitly, we find (with $\tilde {g}_{LV}^{(p)}=2 {g}_{LV}^{u}+{g}_{LV}^{d}$ and $\tilde
{g}_{LV}^{(n)}=2 {g}_{LV}^{d}+{g}_{LV}^{u}$) 
\begin{align}
\tilde {g}_{LV}^{(p)} ={g}_{LV}^{q}/Q_q &= -\frac{\alpha}{2\pi \, q^2}
    \big[ G(m_\mu, m_\mu, m_e, m_h, q^2, Y) - G(m_\mu, m_\mu, m_e, m_h, 0, Y) \nonumber\\
    &\hspace{1.8cm}
      + G(m_\mu, m_e, m_e, m_h, q^2, Y) - G(m_\mu, m_e, m_e, m_h, 0, Y) \big] \,.
    \label{eq:gL}
\end{align}
Here, $Q_q$ is the charge of quark $q$,
and $\tilde {g}_{RV}^{(p)}$,
${g}_{RV}^{q}$ is given by Eq.~\eqref{eq:gL} with the replacement $Y\to
Y^\dagger$. The loop function in \eqref{eq:gL} is
\begin{align}
  \begin{split}
    G(m_i, m_f, m_j, m_h, q^2, Y) &=
      \int_0^1 \! dx\int_0^{1-x}\,dy\, \Big[ Y_{jf} Y_{if}^* \log \Delta
      - \frac{1}{\Delta} m_i m_j z^2 Y_{fj}^* Y_{fi}\\
    & - \frac{1}{\Delta}
               \Big(m_f m_j z Y_{fj}^* Y_{if}^*
              + m_f m_i z Y_{jf} Y_{fi}
              + \big[q^2 x y  + m_f^2 \big] Y_{jf} Y_{if}^*\Big)
    \Big] \,,
    \label{eq:mue-G}
  \end{split}
\end{align}
where we have defined $\Delta\equiv  z m_h^2 - x z \, m_j^2 - y z \, m_i^2 +
(x+y) m_f^2 - x y q^2$ and $z=1-x-y$. Note that we subtract the value of the
one-loop vertex correction at $q^2 = 0$, which gets absorbed into the wave
function and mass renormalizations.  ${g}_{LV}^{q}$ and ${g}_{RV}^{q}$ also
receive two-loop contributions from diagrams similar to the ones relevant for
$c_L$, $c_R$ (see Fig.~\ref{fig:2loop}).  To the best of our knowledge, analytic
expressions for these contributions are not available in the literature, and
are thus not included in our numerical results.
While the one-loop vector contributions are smaller than the one-loop
dipole ones in the $\mu \to e$ conversion rate, and can be neglected, it would
be desirable to also evaluate the two-loop vector terms in order to verify
that all numerically important contributions have been taken into account.

All other Wilson coefficients in Eq.~\eqref{eq:mue:eff} are zero,
$g_{LP}^q=g_{RP}^q=g_{LA}^q=g_{RA}^q=g_{LT}^q=g_{RT}^q=0$.

When computing the $\mu\to e$ conversion rate in nuclei care must be taken
to account for the nuclear matrix elements $\bra{N} \bar{q} q \ket{N}$,
$\bra{N} \bar{q} \gamma^\mu q \ket{N}$ and $\bra{N} F^{\mu\nu} \ket{N}$ and for
the overlap of the initial muon wave function and the final state electron wave
function.  We follow~\cite{Kitano:2002mt} and obtain
\begin{align}
\begin{split}
  \Gamma(\mu \to e \text{ conversion}) =&
    \Big|
      -\frac{e}{16 \pi^2} c_R D + \tilde{g}_{LS}^{(p)} S^{(p)} + \tilde{g}_{LS}^{(n)} S^{(n)}
        +  \tilde{g}_{LV}^{(p)} V^{(p)}
    \Big|^2 \\
    +&
    \Big|
     - \frac{e}{16 \pi^2} c_L D + \tilde{g}_{RS}^{(p)} S^{(p)} + \tilde{g}_{RS}^{(n)} S^{(n)}
        +  \tilde{g}_{RV}^{(p)} V^{(p)}
    \Big|^2 \,.
  \label{eq:mu-e-conversion}
  \end{split}
\end{align}
The electromagnetic penguin and vector contributions were already given above.
Note that vector couplings to neutrons are absent due to the neutron's
vanishing electric charge.  The scalar coefficients for proton and neutron
coupling are given in terms of the quark level coefficients by
\begin{align}
  \tilde{g}_{LS,RS}^{(p)} =  \sum_{q}  g_{LS,RS}^q \,
    \frac{m_p}{m_q} f^{(q,p)} \,,\qquad   \tilde{g}_{LS,RS}^{(n)} =  \sum_{q}  g_{LS,RS}^q \,
    \frac{m_n}{m_q} f^{(q,n)}\,.
  \label{eq:gLSp}
\end{align}
Here, the sum runs over all quark flavors, $q=u,d,s,c,b,t$.

The nucleon matrix elements $f^{(q,p)} \equiv \bra{p} m_q \bar{q} q \ket{p} /
m_p$ are calculated according to \cite{Ellis:2008hf}, but using an updated
value for the nucleon sigma term $\Sigma_{\pi N} = 55$~MeV~\cite{Young:2009ps} (If the value
of the nucleon sigma term is even smaller, as indicated by recent unquenched lattice results, our bounds would become weaker).
The nucleon matrix elements are numerically
\begin{align}
  f^{(u,p)} =  f^{(d,n)}=  0.024 \,, \qquad
  f^{(d,p)} = f^{(u,n)} = 0.033 \,, \qquad
  f^{(s,p)} =  f^{(s,n)} =  0.25  \,,
\end{align}
while the contributions from the heavier quarks are
\begin{align}
  f^{(c,p)} = f^{(b,p)} = f^{(t,p)} = \frac{2}{27} \Big( 1 - \sum_{q=u,d,s} f^{(q,p)} \Big) \,.
\end{align}
with the same values for neutrons.  In the above expressions, $m_q$ denotes a
quark mass, $m_p$ is the proton mass, and $m_n$ is the neutron mass.  The
coefficients $D$, $V^{(p)}$, $S^{(p)}$, and $S^{(n)}$ are overlap integrals of
the muon, electron and nuclear wave function. They are tabulated for various
target materials in~\cite{Kitano:2002mt}. The best limits are obtained from
bounds on $\mu\to e$ conversion on gold, $\Gamma(\mu \to e)_{\rm
Au}/\Gamma_{\rm capture~Au}< 7 \times 10^{-13}$ ($90\%$ CL)
\cite{Bertl:2006up}, for which in units of $m_\mu^{5/2}$ the overlap integrals
are $D=0.189$, $S^{(p)}=0.0614$, $V^{(p)}=0.0974$, $S^{(n)}=0.0918$,
using the same distributions for neutrons and protons in the
nucleus. For the SM capture rate, we use a value $\Gamma_{\rm
capture~Au}=13.07\times 10^6$~s$^{-1}$ in the calculation~\cite{Kitano:2002mt}.

\bibliographystyle{apsrev}
\bibliography{./higgs-fv}

\end{document}